\documentclass[journal]{IEEEtran}
\usepackage{setspace,color,comment,graphicx,epstopdf}
\pdfoutput=1
\usepackage{amsmath}
\usepackage{algorithmic}
\usepackage{graphicx}
\usepackage{subfigure}
\usepackage{cite}
\usepackage{algorithmic}
\usepackage{algorithm}
\usepackage{epsfig}
\usepackage{multirow}
\usepackage{multicol}

\begin{document}
\title{Loss-resilient Coding of Texture and Depth for Free-viewpoint Video Conferencing}

\author{Bruno Macchiavello,
        Camilo Dorea,
        Edson M. Hung,
        Gene Cheung~\IEEEmembership{Senior Member,~IEEE}
        and Wai-tian Tan~\IEEEmembership{Senior Member,~IEEE}% <-this % stops a space
\thanks{Bruno Machiavello,~~Camilo Dorea,~~Edson M. Hung are with the
University of Brasilia - Asa Norte,~~CEP: 70910-900,~~Brasilia - DF,~~Brazil \newline
Email Address: \texttt{\{bruno, camilo, mintsu\}@image.unb.br}.}% <-this % stops a space

\thanks{Gene Cheung is with the National Institute of Informatics, 2-1-2, Hitotsubashi, Chiyoda-ku,
Tokyo, 101--8430, Japan \newline
Email Address: \texttt{cheung@nii.ac.jp}.}

\thanks{Wai-tian Tan is with the Hewlett-Packard Laboratories, 1501 Page Mill Road, Palo Alto, CA 94304, USA \newline
Email Address: \texttt{wai-tian.tan@hp.com}.}}

\maketitle

\begin{abstract}
Free-viewpoint video conferencing allows a participant to observe the 
remote 3D scene from any freely chosen viewpoint.
An intermediate virtual viewpoint image is commonly synthesized using two 
pairs of transmitted texture and depth maps from two neighboring captured 
viewpoints via depth-image-based rendering (DIBR).
To maintain high quality of synthesized images, 
it is imperative to contain the adverse effects of network packet losses 
that may arise during texture and depth video transmission.
Towards this end, we develop an integrated approach that exploits 
the representation redundancy inherent in the multiple streamed videos---a 
voxel in the 3D scene visible to two captured views is sampled and
coded twice in the two views. In particular, at the receiver we first develop 
an error concealment strategy that adaptively blends corresponding pixels 
in the two captured views during DIBR, so that pixels from the more 
reliable transmitted view are weighted more heavily. We then couple it 
with a sender-side optimization of reference picture selection (RPS) 
during real-time video coding, so that
%the likelihood of long error 
%propagation---typical in differentially coded video---is minimized 
%for blocks containing pixels vital to adaptive rendering at receiver. 
blocks containing samples of voxels that are visible in both views
are more error-resiliently coded in one view only, given adaptive blending
will erase errors in the other view.
Further, synthesized view distortion sensitivities to texture versus
depth errors are analyzed, so that relative importance of texture
and depth code blocks can be computed for system-wide RPS optimization. 
%As a result, for portions of the 3D scene that are observable from both 
%captured views, our system can judiciously manipulate data redundancy 
%by providing preferential protection to only one view for effective
%adaptive blending at receiver.
Experimental results show that the proposed scheme can outperform 
the use of a traditional feedback channel by up to $0.82$ dB on average 
at 8\% packet loss rate, and by as much as $3$ dB for particular frames.
\end{abstract}


% \fi
\IEEEpeerreviewmaketitle

\section{Introduction}

\IEEEPARstart{T}{he} 
demand for ever improving quality of video communication has already
made high definition video conferencing a basic feature not
only on computers, but smart-phones as well.
A free-viewpoint video conferencing system~\cite{kubota07} that continuously 
alters the displayed images according to a user's real-time selected 
viewpoints---e.g., through \textit{motion parallax}~\cite{zhang09}, where
detected head movements trigger corresponding rendering of images as 
viewed from the observer's viewpoint---can enhance an observer's depth 
perception in the 3D scene, and bring video communication to a new level 
of immersion and realism. The techniques for capturing~\cite{fujii06}, 
representing~\cite{flierl07, merkle07a}, and 
synthesizing~\cite{Tanimoto08, tian09} free-viewpoint video have been 
well studied.
Instead, in this paper we address the problem of robust low-latency 
streaming of free-viewpoint video over packet-loss prone networks.

%free-viewpoint is 
%a desirable degree of freedom for observers, which 
%desirable and
%can improve human's 
%visual perception of depth in the 3D scene through \textit{motion parallax}. 
We adopt the widely employed depth-image-based rendering (DIBR) 
approach~\cite{tian09} to provide free viewpoint.
%is an image synthesis technique that 
%enables rendering of an image from an arbitrarily chosen virtual viewpoint 
%in a multiview video system. It 
As the name suggests, in addition to typical RGB images (texture),
DIBR requires also depth images (per-pixel physical distances between 
captured objects in the 3D scene and capturing camera) to synthesize
a freely chosen viewpoint image.  
Depth maps can be obtained by estimation algorithms like stereo-matching, 
or depth sensors like time-of-flight cameras~\cite{gokturk04}. 
While recent proposals such as \cite{maugey12, daribo12} suggest 
transmission of one pair of texture and depth maps from a single 
camera-captured viewpoint for synthesis of a defined neighborhood of 
viewpoints, this usually leads to larger disoccluded 
regions\footnote{Disoccluded region is a spatial region in the synthesized
view with no corresponding pixels in the reference views during DIBR. This
issue is discussed in details in Section~\ref{sec:conceal}.} 
in the synthesized view that require image inpainting~\cite{oh09, daribo10}
to complete the image, resulting in high complexity.
%and the additional neighboring view required for stereo vision is 
%rendered~\cite{fehn04}.   
We hence assume the more customary approach of transmitting 
texture and depth maps of two neighboring captured
views~\cite{kurutepe07, zhang09} to ensure quality reconstruction 
of freely chosen intermediate virtual views.

The transport of texture and depth videos for multiple views presents 
both challenges and opportunities.
On one hand, the complex dependency of rendered view quality on
both depth and texture maps of captured views needs to be adequately modeled
to realize effective error-resilience measures.  On the other hand, 
the inherent representation redundancy in multiple texture-plus-depth 
videos---\textit{a voxel\footnote{A voxel is a volume element 
representing a value on a regular grid in 3D space~\cite{voxel}.} 
in the 3D scene visible to two captured views is 
sampled and coded twice in the two views}---suggests that errors 
in one view can be ameliorated by correct delivery of another view. 
Towards this end, we develop an integrated approach in this paper where
receiver and sender are jointly designed to exploit and 
manipulate the inherent representation redundancy present in depth and 
texture videos from multiple views. To the best of our knowledge,
this is the first work in the literature that exploits representation
redundancy in multiple texture-plus-depth videos for loss-resilient
streaming.

To exploit representation redundancy at the receiver, 
we first develop an error model that tracks the 
reliability of each code block in each transmitted view given
observed packet loss events. During DIBR-based rendering of a virtual
view, where a synthesized pixel is
typically computed as a convex combination of the two corresponding pixels
in the two captured views, we perform \textit{adaptive blending}, 
so that the corresponding pixel from the more reliable transmitted
view is assigned a heavier weight. 
Adaptive blending uniquely targets quality improvement of the DIBR 
synthesized view, and is thus complementary to conventional error 
concealment methods that improve reconstruction quality of individual 
captured views~\cite{chen08, yeo09}. 

To exploit representation redundancy at the sender, we design 
an optimization of reference picture selection (RPS) 
during real-time video coding, so that blocks containing samples of 
voxels that are visible in both views are more error-resiliently coded 
(e.g., through intra-coding or using a reference block in a previous 
frame that has been acknowledged) in one view only. This is possible 
because aforementioned adaptive blending is performed at decoder, 
suppressing errors in one view if corresponding blocks in the other 
view are correctly delivered. 
The expected reconstructed error of a block predicted from a given past 
frame is computed via a set of recursive equations we derive with
computation efficiency in mind.

Finally, because depth maps are auxiliary information that only aid in
the construction of the synthesized views but are not themselves directly
observed, the synthesized view distortion sensitivities to errors
in depth maps are clearly different to errors in texture maps. 
To guide preferential protection, we analyze synthesized view 
distortion sensitivities to texture versus depth errors, so that 
relative importance of texture and depth code blocks can be computed 
for system-wide RPS optimization. 

We summarize our texture-plus-depth video streaming optimization 
for free-viewpoint conferencing as follows:
\begin{enumerate}
\item At receiver, adaptive blending is performed so that the corresponding
pixels from the more reliable transmitted view are weighted more heavily
during DIBR-based view synthesis.
\item At sender, we perform per-block RPS, so that the expected 
error, stemming from error propagation of a lost reference block, computed 
using a set of computation-efficient recursive equations, is minimized for 
blocks containing important pixels vital to adaptive blending at receiver.
\item Synthesized view distortion sensitivities to errors in
texture versus depth maps are analyzed, so that preferential RPS 
for coding of texture and depth blocks can be optimized.
\end{enumerate}
Experimental results show that the proposed scheme can outperform 
the use of a traditional feedback channel by up to $0.82$ dB on average 
at 8\% packet loss rate, and by as much as $3$ dB for particular frames.

The rest of the paper is organized as follows. We first discuss related 
work in Section~\ref{sec:related}. We then overview our streaming system
in Section~\ref{sec:system}. We discuss our receiver error concealment
strategy using adaptive blending in Section~\ref{sec:conceal}. We discuss
our sender RPS optimization in Section~\ref{sec:formulate}. In 
Section~\ref{sec:complexity}, an analysis of the added computational 
load of our proposed algorithms is presented. Finally, 
experimental results and conclusions are presented in 
Section~\ref{sec:results} and \ref{sec:conclude}, respectively.

\section{Related Work}
\label{sec:related}

We divide the discussion of related work into four sections. We first
discuss related work in multiview and free-viewpoint video coding. 
We next discuss depth map denoising work in the compression and 
view synthesis literature.
We then discuss related work in error-resilient video streaming
and error concealment for conventional 2D video. Finally, we juxtapose
the contributions of this paper to our own earlier work on the same
topic.

\subsection{Multiview and Free Viewpoint Video Coding}
\label{subsec:mvc}

Multiview video coding (MVC) is concerned with the compression of 
texture videos captured from multiple nearby viewpoints. Early works in 
MVC~\cite{flierl07, merkle07} focused on exploiting signal redundancy
across view using \textit{disparity compensation} for
coding gain---matching of code blocks between neighboring view pictures
for efficient signal prediction. Best-matched block in the reference
frame is identified by a \textit{disparity vector}, similar to
motion vector in motion compensation used in conventional 2D video
coding standards like H.263~\cite{h263} and H.264~\cite{wiegand03}.
However, given temporal redundancy has already been exploited via 
motion compensation, and neighboring temporal frames tend to be
more similar than neighboring inter-view frames due to typical
high-frame-rate videos captured by cameras, it was shown that 
additional coding gain afforded by disparity compensation is noticeable
but not dramatic (about 1dB difference in PSNR~\cite{merkle07}). 
Given our goal is loss-resilient video streaming, to avoid potential 
inter-view error propagation in disparity compensated multiview video, 
we perform motion estimation / compensation independently in each view. 
(We note further that having independent encoders for different 
views fits the \textit{multiterminal video coding} 
paradigm~\cite{zixiang11}, which, besides benefits of lower overall 
encoding complexity, holds promises of better compression performance in 
the future via distributed source coding theory.)

While MVC offers view-switches at receiver only among the discrete
set of captured and coded views, transmitting both texture and 
depth videos of nearby views---known as ``texture-plus-depth" 
format~\cite{merkle07a}---enables the observer to choose any 
intermediate virtual view for DIBR-based image rendering and display.
Given compression of texture maps has been well studied in the past
decades (e.g., coding standards such as H.263~\cite{h263} and 
H.264~\cite{wiegand03}), 
much recent works thus focused specifically on depth map 
compression.
%~\cite{kim09, kim10, cheung10pcs, shen10pcs, cheung11icip, 
%gautier12, valenzise12, hu12}.
These works can be divided into three classes. The first 
class~\cite{shen10pcs, gautier12, hu12} designed
specific coding tools (e.g., graph-based transform (GBT) 
\cite{shen10pcs, hu12}) that tailored to the unique signal 
characteristics of depth maps, such as sharp edges and smooth interior 
surfaces. In our streaming system, for simplicity we assume H.264 is
employed for standard-compliant compression of texture and depth videos
in each captured view. However, we note that new coding tools such as 
GBT can be incorporated into our streaming optimization easily, if 
standard-compliant solution is not required. 

The second class~\cite{kim09, kim10, cheung10pcs, cheung11icip, valenzise12} 
observed that a depth map is a source of 
auxiliary information that assists in the synthesized view construction 
at decoder, but is itself never directly observed. Thus, one can design 
coding optimizations that minimize the indirect synthesized view 
distortion rather than the direct depth map distortion.
Our analysis of synthesized view distortion sensitivities to
errors in texture and depth maps is an extension of \cite{cheung11icip}
from coding optimization to loss-resilient streaming optimization.

The third class~\cite{merkle12, daribo12_pcs} exploited 
correlations between texture and depth maps of the same view (such 
as common edge patterns) for compression gain. In our previous 
study~\cite{bruno12-2}, we have concluded that an important depth
block (e.g., one that contains an edge between a foreground object
and background) can be more critical to synthesized view quality than a 
texture block. Predicting depth blocks from coresponding texture blocks 
(as done in \cite{merkle12} as an extension of HEVC video coding standard
to texture-plus-depth format)
would entail unnatural dependency of a more important depth block on 
a less important texture block. Given depth video typically requires 
roughly 10\% of the total bitrate and the goal is loss resiliency
of free viewpoint video streaming, we choose to forego inter-component
prediction and code texture and depth videos of the same view 
independently using H.264.

\subsection{Depth Map Denoising}

It is observed in the depth map compression and view synthesis literature
that multiple depth maps across different views may
not be consistent due to acquisition errors and/or 
compression artifacts. Thus, to eliminate inter-view inconsistency,
one can perform pre-processing of depth maps at encoder before
compression for coding gain~\cite{li12}, or post-processing at decoder
after decoding for improvement in view synthesis 
quality~\cite{furihata10, helgason12}. In general, the cause of the 
inconsistency cannot be determined during processing, 
however, and so when left and right depth maps are in conflict, 
one cannot determine which one (or both)
should be corrected for inter-view consistency.
In contrast, in our free-viewpoint video streaming application,
the decoder can track the packet loss events to determine the
reliability of the left map versus the right map (see
Section~\ref{sec:conceal} for details). Thus, during DIBR view synthesis
an adaptive blending procedure can be applied to assign heavier weight 
to the corresponding pixel that is more reliable.

\subsection{Error-resilient Video Streaming}

The problem of error control and concealment has been actively studied for
many years, and an overview of general approaches can be found 
in~\cite{wang98}.
One particular effective mechanism to control error propagation for systems 
with live encoder is reference picture selection (RPS). 
In \textit{reactive} RPS, a live encoder reacts to receiver 
feedbacks by avoiding the use of notified loss-affected past frames as 
reference for coding of future frames.  In 
\textit{proactive} RPS, long prediction chains are avoided during
video encoding without incorporating real-time client feedback 
information. 
Both reactive and proactive RPS can be implemented to be 
compatible with video syntax of H.263 version 2, MPEG-4 and H.264.
Reactive RPS incurs higher bit overhead only when needed, but
suffers error propagation of up to one round-trip time (RTT) at decoder.
Proactive RPS incurs an overhead regardless of whether there are 
actually losses, and should only be applied preferentially to more 
important parts of the video.
A study of adaptively choosing different RPS approaches in streaming of 
single-view H.264 video is given in~\cite{Wang_RPS_09}.
In contrast, we employ at sender proactive RPS with 
feedbacks~\cite{cheung07transcsvt} (call \textit{proactive-feedback} in the 
sequel) for selected code blocks in the depth and texture sequences during 
real-time video encoding, so that the likelihood of long error 
propagation---typical in differentially coded video---is 
minimized for blocks containing pixels vital to adaptive rendering at 
receiver. Our proactive-feedback approach integrates feedback 
by computing estimated distortion for all frames with and without known 
loss information.

%Unequal error protection is not a new idea for single-view video. In this
%paper, we examine how this could be performed when transmitting multiple 
%views of depth and texture using an optimization framework that takes 
%into account concealment actions.

There are many other techniques developed for improving resilience of 
single-view video, including error concealment, retransmissions and use of 
error correcting codes.
In addition to commonly employed Reed-Solomon codes, there are also recent 
low-delay ``streaming codes'' that allows fast recovery of earlier 
transmitted packets without requiring all losses to be 
corrected~\cite{Martinian07}.
Modern video compression formats also have syntax support for error resilience.
For example, flexible macroblock ordering~\cite{Lambert06} is an effective 
error resilient tool in H.264 that can avoid loss of large contiguous 
regions to make error concealment more effective.  Another 
example is data partitioning~\cite{Stockhammer04},
which allows decomposition of compressed single-view video into layers of
different importance for preferential protection, and has been implemented in
Google's desktop Chrome browser as part of the WebRTC stack.
Furthermore, it is possible to explicitly transmit addition data in various 
forms to aid error concealment~\cite{Troger_11}.

Many of these techniques are applicable to free-viewpoint conferencing and are
complementary to multi-view error concealment methods as well as to adaptive
error-resilient source coding methods. To limit the scope of this paper, 
however, we do not explicitly consider special methods to conceal individual 
transmitted views nor do we employ retransmissions or error correction codes.

Joint optimization of compression, concealment, and channel loss has also been
considered.  For example, the work in~\cite{eisenberg2003}, considers a formulation
that allocates transmission power to jointly optimize the average and variance of 
distortion for single view video.
Finally, though there have been streaming optimizations proposed in the
literature for stereoscopic video~\cite{tan09} and multiview 
video~\cite{hou10}, to the best of our knowledge we are the first to
propose loss-resilient coding specifically for interactive 
free-viewpoint video where texture and depth maps of two neighboring camera
viewpoints are transmitted~\cite{xiu12}. In particular, leveraging on 
the observation that the representation of multiple texture-plus-depth is  
inherent redundant, we design adaptive blending for view synthesis
at decoder and optimized RPS coding scheme at encoder for optimal
end-to-end streaming performance.

\subsection{Contribution over Our Previous Work}

In \cite{bruno12}, a similar scheme to minimize expected synthesized
view distortion based on selection of reference frame at the block level 
was proposed for depth maps only. 
%There, a method was presented to determine the importance of a MB in a 
%depth map frame, 
%as the curvature parameter ($a$) of a quadratic function that fits the 
%error function, i.e., the difference between the incorrectly mapped pixel 
%and the correct pixel value due to a disparity error. 
In another previous work \cite{bruno12-2}, we extended the idea in
\cite{bruno12} to encoding of both texture \textit{and} depth maps. We also
expanded the coding modes available to each MB to include 
intra block coding. Furthermore, we proposed an optimization algorithm for 
encoding of both texture and 
depth maps that nonetheless remains computationally efficient. 
We note that the recursive model to estimate distortion of a synthesized 
block using both texture and depth information is inspired by 
derivations proposed in \cite{zhou11tcsvt}.
We have also studied optimized strategy for motion compensated video in 
an RPS framework for single-view video at the frame level 
\cite{cheung07transcsvt}. 

In this work, we make significant contributions over our earlier works
\textit{by exploiting the inherent representation redundancy in multiple 
texture-plus-depth videos---a voxel in the 3D scene visible to two 
captured views is sampled and coded twice in the two views}. 
Specifically, given the intermediate 
virtual view is obtained by interpolating texture of two neighboring 
captured views, an error concealment strategy can be performed during 
view synthesis where the corresponding pixel with higher reliability
is weighted more heavily during pixel blending. 
Correspondingly, the encoder can perform an optimization algorithm so that 
pixels that are visible from both captured views
can be unequally protected: only pixels from one view are heavily protected
to ensure good synthesized pixels at decoder. 
Note that in this newly updated encoding algorithm, the expected 
synthesized view distortion depends on the expected error from both coded 
views. However, we design a novel problem formulation, so that there is 
no inter-dependency between frames of the same time instant in both views, 
easing the system-wide RPS optimization. 
%This new encoding mode is applied to both texture and depth maps, and 
%the modified view synthesis process takes into consideration errors 
%in all the input signals.  

%This paper extends to multiple-views, with
%In contrast, we 
%optimize RPS at the MB level according to the importance of each MB 
%in texture and depth maps, in order to minimize the 
%expected synthesized view distortion. 

\section{Real-time Free Viewpoint Video Streaming System}
\label{sec:system}

\begin{figure}[ht]
\begin{minipage}[b]{1.0\linewidth}
  \centering
  \includegraphics[width=8cm]{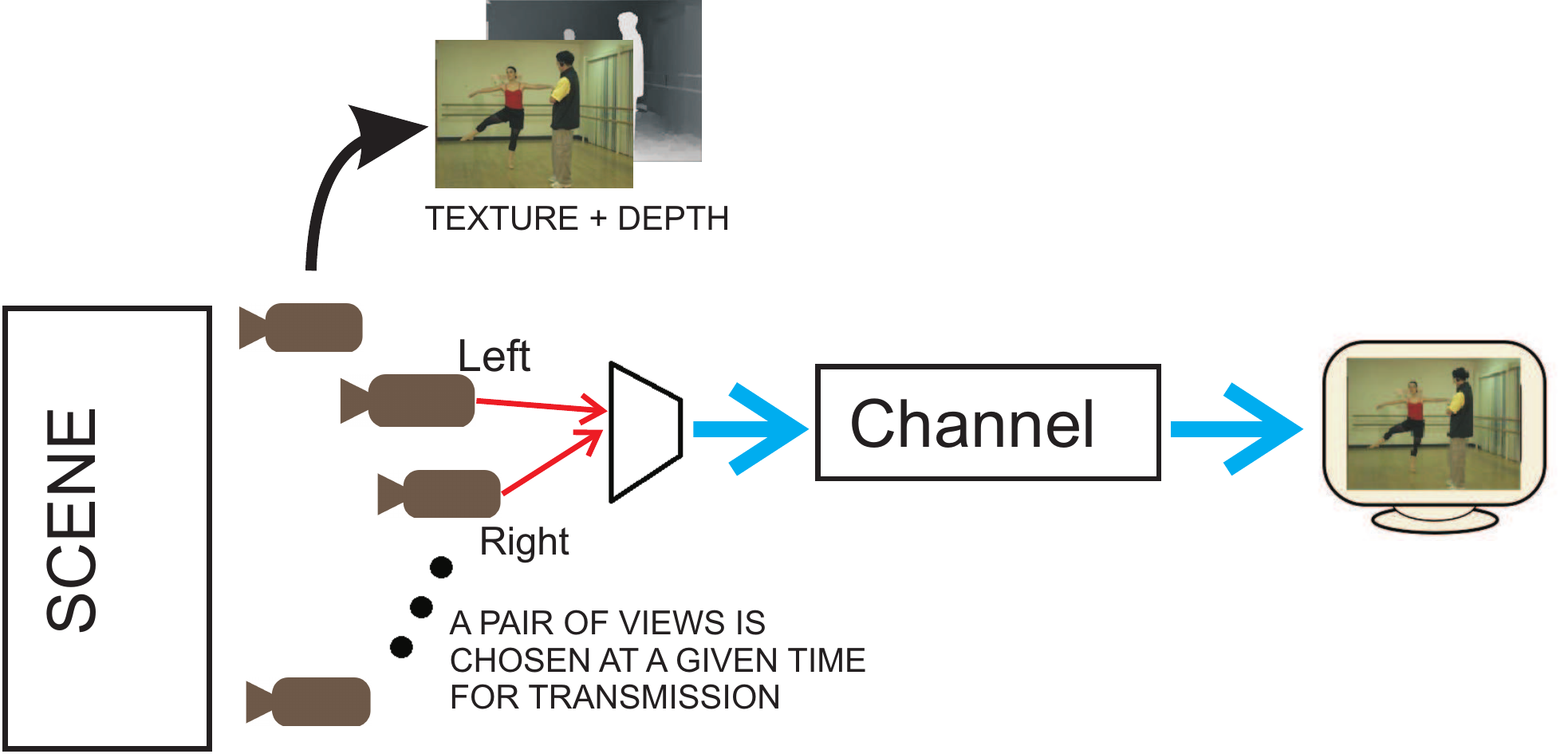}
  %\centerline{\psfig{figure=mv_video_system.eps,width=8.8cm} } 
\end{minipage}
\caption{A bandwidth-efficient free-viewpoint video streaming system 
dynamically selects two views for transmission.}
\label{fig:mv_system}
\end{figure}

We overview our proposed real-time free viewpoint video streaming system
in this section. At sender, we assume a large array of closely
spaced cameras are capturing videos of a 3D scene simultaneously.
See Fig.~\ref{fig:mv_system} for an illustration. Specific constraints are commonly
imposed upon scene acquisition in order to support this application as well as
other 3D video services. Cameras are deployed in a 1D parallel arrangement with narrow acquisition angles and a baseline between cameras of the order of 5 cm \cite{MPEG3Dcall}. Rectification, performed prior to encoding, is often necessary to eliminate misalignments and provide views in linear and parallel arrangement. These conditions guarantee that disparities between views
are limited to 1D shifts along the x-axis. Naturally, accuracy in camera calibration parameters is also required. Furthermore, color consistency among cameras should be assured upon capturing and/or enforced through image processing methods \cite{tanimoto11} prior to encoding.

%Color consistency among the multiple views should be adjusted as accurate as possible prior to capturing. The multiple cameras should be white balanced (gamma correction). Most modern cameras provide functionalities for this. 

Besides texture maps (e.g., color images like RGB), we assume
depth maps (per-pixel physical distances between capturing camera and 
the captured objects in the 3D scene) of the same resolution and 
from the same camera viewpoints as the texture maps can also be obtained, 
either via stereo-matching algorithms using neighboring texture maps, 
or captured directly using depth-sensing cameras~\cite{gokturk04}. 
The assumption that both texture and depth maps are available
from the same camera viewpoints is a common one for 
3D visual data represented in the now popular texture-plus-depth 
format~\cite{merkle07a}. Armed with both texture and depth maps,
intermediate virtual views can be synthesized via a 
depth-image-based rendering (DIBR) technique like
3D warping~\cite{mark97}, enabling user to select and render
image at any desired viewpoint for observation. The ability to choose 
any viewpoint for image rendering is called 
\textit{free viewpoint}~\cite{tanimoto11}.

The number of cameras capturing video at the same time 
can be quite large---up to 100 cameras were used in \cite{fujii06}.
Hence real-time encoding and transmitting all of them from sender
to receiver would translate to too large of a network cost.
Instead, we assume here that the current view $v$ that the receiver
is observing is constantly fed back to the sender. The sender
can then estimate the range of virtual views the receiver will 
choose and observe during the next round-trip time (RTT), and then
select only \textit{two} neighboring camera views that can enable
rendering of those virtual views for real-time
coding and transmission of the corresponding texture and depth maps.
This idea of selecting only a subset of camera captured views 
for efficient network transmission is also exploited in
\cite{kurutepe07}. In this paper, we focus on the optimal transmission
and error concealment of the texture and depth maps from
those two selected views.

\section{Error Concealment Strategy for DIBR-based Synthesis}
\label{sec:conceal}
% \subsection{Error Concealment for DIBR View Synthesis System}

We first review a common DIBR view synthesis procedure that assumes 
no information loss in either depth or texture maps.  We then derive 
formulas for estimates of texture and disparity error 
for every macroblock (MB) due to packet losses in %closed-loop 
motion-compensated video. 
Finally, we discuss how an error concealment 
strategy for the synthesized view---\textit{adaptive blending}---can 
be performed given the estimated texture and disparity errors.

\subsection{Standard Loss-free View Synthesis}
\label{subsec:standardVS}

In a multiview DIBR system, an intermediate virtual view %, $0 \leq v \leq 1$, 
is interpolated using texture and depth maps of two neighboring captured 
views %, $0$ and $1$, 
via a DIBR technique like 3D warping~\cite{mark97}. We focus on the fundamental view
blending step upon which our proposal is based. Note that prior to blending, mechanisms aimed at robustness, such as consistency checks, may condition the correspondences among views. 
The reader is referred to \cite{tian09} for a complete description of view synthesis procedures.
%Reply R1 C3

%various pre- and post-processing steps, such as described in \cite{tian09}, were used in our implementation but are not addressed here. %Reply R1 C3

As a convention, we label left and right views as 0 and 1, respectively. 
Given the texture map of the left view $X^0_t$ and right view $X^1_t$ at 
time $t$, the pixel value $S^v_t(i,j)$ at coordinate $(i,j)$ of a 
synthesized view $v$ depends on whether we can find a corresponding pixel 
$(i, j^0)$ from the left view and a corresponding pixel $(i, j^1)$ from 
the right view.
Specifically, the synthesized image of view $v$, $0 \leq v \leq 1$, 
has pixel value $S^v_t(i,j)$ at coordinate $(i,j)$ given by:

\small
\begin{equation}
S^v_t(i,j) = \left\{\hspace{-1.5mm} \begin{array}{ll} 
(1-v)  X_t^0(i, j^0) + v  X_t^1(i,j^1) 
& \mbox{if L \& R pixels exist} \\
X_t^0(i, j^0) & \mbox{if only L pixel exists} \\
X_t^1(i,j^1) & \mbox{if only R pixel exists} \\
\mbox{hole} & \mbox{o.w.}
\end{array} \right.
\label{eq:DIBR}
\end{equation}
\normalsize

\noindent where the weights $1-v$ and $v$ are inversely proportional to the 
virtual view's distance from view 0 and 1. A chosen inpainting algorithm like~\cite{oh09, daribo10} is then applied to fill in
a small number of hole pixels that have no correspondence in both views 0 and 1. 
In other words, weighted blending is performed if two correspondences are 
found, while single-pixel mapping and inpainting are invoked if one or zero 
correspondence is found, respectively. See \cite{tian09} for details
of a standard DIBR view synthesis implementation.

A corresponding pixel $(i,j^0)$ in the left texture map $X_t^0$ is a
pixel that, given associated disparity $Y_t^0(i,j^0)$ between left and right views subject to normalization,   %reply R1 C4
shifts horizontally
by $Y_t^0(i,j^0) * v * \eta$ pixels to synthesized pixel of coordinate 
$(i,j)$ in virtual view $S_t^v$, where $\eta$ is a scaling factor 
determined by the distance between the capturing cameras. We can write 
$j^0$ and $j^1$ for the two corresponding pixels in left and right 
texture maps as:
\begin{eqnarray}
j & = & j^0 - Y_t^0(i,j^0) * v * \eta \nonumber \\
j & = & j^1 + Y_t^1(i,j^1) * (1-v) * \eta
\label{eq:correspond}
\end{eqnarray}

One interpretation of (\ref{eq:DIBR}) is that the same physical 
point in the 3D scene (called \textit{voxel} in computer graphics
literature~\cite{shum07}) has observed intensities $X^0_t(i,j^0)$ 
and $X^1_t(i,j^1)$ from views $0$ and $1$, respectively. 
Hence, the intensity of the same voxel at an intermediate 
view can be modeled as a convex combination of the two 
corresponding values. 
If the surface of the observed physical object has
\textit{Lambertian reflectance}~\cite{shum07}, then a voxel 
will have (roughly) the same intensity from different viewpoints; 
i.e., $X^0_t(i,j^0) \approx X^1_t(i,j^1)$. In such case of 
\textit{redundant representation} (same intensity value is recorded in
two corresponding coordinates at left and right maps), synthesized pixel 
$S^v_t(i,j)$ can be perfectly reconstructed even if one of the two 
corresponding pixels, $X^0_t(i,j^0)$ and $X^1_t(i,j^1)$, is corrupted by 
channel noise, assuming the decoder knows which corresponding pixel is 
erred. Based on this observation, and assuming that the majority
of synthesized pixels are of objects with Lambertian reflectance 
surfaces\footnote{While common objects like glass and mirrors do not have
Lambertian surfaces, if the two captured viewpoints are sufficiently
close, voxels of these objects can nonetheless be approximated as 
having the same intensity values from different but nearby viewpoints.}, 
we develop the following error concealment strategy during view synthesis.

The key idea of the error concealment strategy for DIBR view synthesis is 
as follows. When there are two corresponding pixels $X^0_t(i,j^0)$ 
and $X^1_t(i,j^1)$ for a given synthesized pixel $S^v_t(i,j)$, 
decoder can choose to \textit{reweigh} the combination of the pixels
depending on the reliability of the two reconstructed pixels. 
For example, decoder can use only one corresponding pixel
for synthesis if the other corresponding pixel is deemed 
totally unreliable. 
%Obviously, this re-weighting strategy is sensible
%only if the surface reflectance is roughly Lambertian. 
We next discuss how we estimate the reliability
of the two corresponding texture pixels $X^0_t(i,j^0)$ and $X^1_t(i,j^1)$
at decoder given observed packet losses in the transmitted maps.

\subsection{Estimating Texture Error at Decoder}
\label{subsec:estTexErr}

We first compute the reliability of the two corresponding texture
pixels $X^0_t(i,j^0)$ and $X^1_t(i,j^1)$, assuming the corresponding
disparity pixels $Y^0_t(i,j^0)$ and $Y^1_t(i,j^1)$ are correct.
Let $m = b(i,j^0)$ be the MB that contains texture
pixel $(i,j^0)$, and let $e_{t,m}$ be the texture error due to packet losses %reply R1 C5
in MB $m$ given differential coding of texture maps.  
Depending on whether MB $m$ is received correctly or not, $e_{t,m}$ will
result in error $e^+_{t,m}$ and $e^-_{t,m}$, respectively:
\begin{equation}
e_{t,m} = \left\{ \begin{array}{ll}
e^+_{t,m} & \mbox{if MB} ~m ~\mbox{is correctly received} \\
e^-_{t,m} & \mbox{o.w.}
\end{array} \right.
\label{eq:decoderE}
\end{equation}

\begin{figure}%[htb]

%\begin{minipage}[b]{.48\linewidth}
%  \centering
% \centerline{\epsfig{figure=figures/modeSelect.eps,width=5.5cm}}
%  \vspace{-0.0in}
% \centerline{\small{(a) blocks along contour}}\medskip
%\end{minipage}
%\hfill
%\begin{minipage}[b]{0.48\linewidth}
  \centering
  \includegraphics[width=9.0cm]{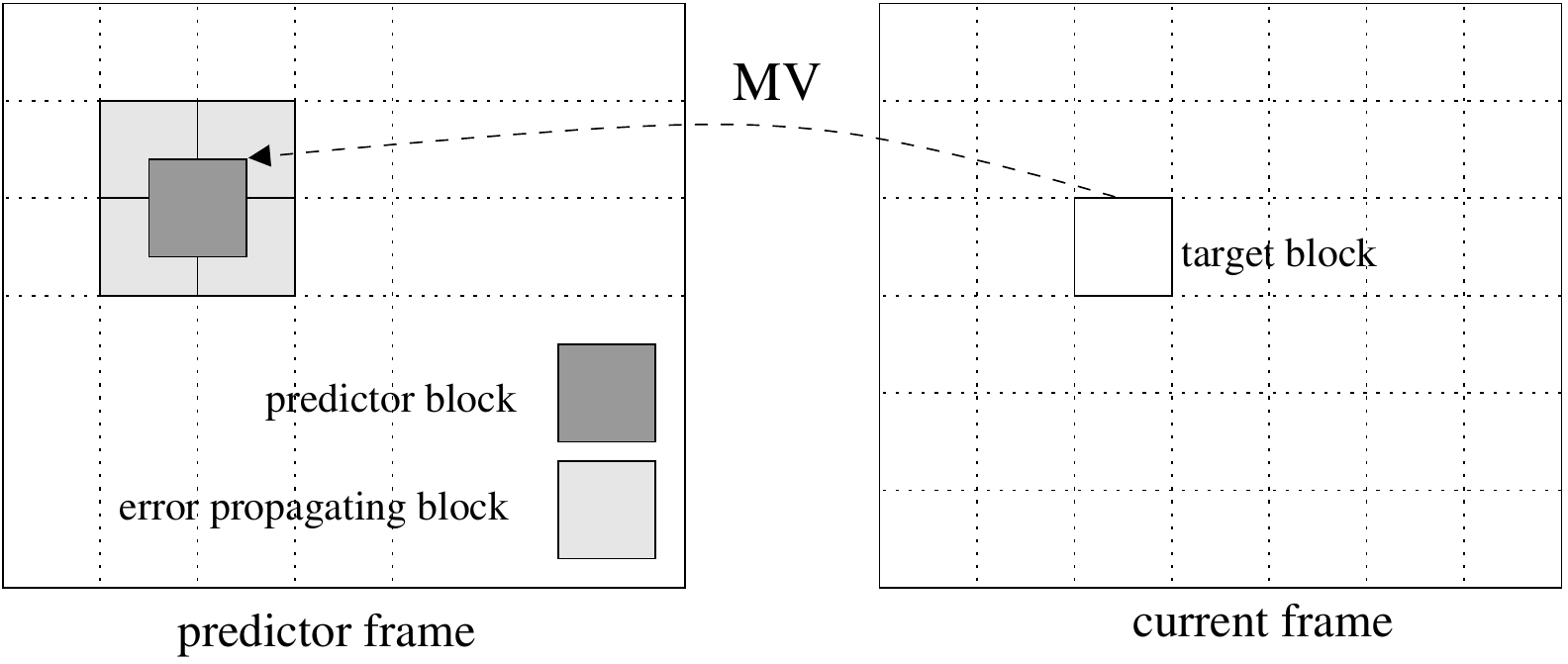}
  %\centerline{\epsfig{figure=motionComp.eps,width=9.0cm}}
  \vspace{-0.0in}
% \centerline{\small{(b) trellis-based mode selection}}\medskip
%\end{minipage}

\caption{Motion prediction in differentially coded video causes
error propagations from predictor block to target block.}
\label{fig:motionComp}

\end{figure}

Consider first $e^+_{t,m}$. If MB $m$ is an intra-coded block
(independently coded) and is correctly received at decoder, then 
$e^+_{t,m} = 0$. 
If MB $m$ is an inter-coded block (differentially coded using a block of a 
previous frame as predictor) pointing to a previous frame $\tau_{t,m}$
with motion vector (MV) $v_{t,m}$, then $e^+_{t,m}$ depends on the errors
of predictor block in frame $\tau_{t,m}$. Note that because the
predictor block indicated by MV $v_{t,m}$ can be located among several 
coded MBs in frame $\tau_{t,m}$, there are typically multiple 
\textit{error propagating} MBs, $k \in v_{t,m}$, that can potentially 
contribute to $e^+_{t,m}$. See Fig.~\ref{fig:motionComp} for an 
illustration. Thus we write $e^+_{t,m}$ recursively as 
follows\footnote{Though we have already argued that inter-view prediction
is not a sensible coding option for our loss-resilient free viewpoint
video streaming framework, (\ref{eq:decoderEplus}) can nonetheless
be modified easily to account for inter-view dependencies if deployed.}:
\begin{equation}
e^+_{t,m}(\tau_{t,m}, v_{t,m}) = \left\{ \begin{array}{ll}
0 & \mbox{\small{if MB $m$ is intra}} \\
\gamma \sum_{k \in v_{t,m}} \alpha_k ~ e_{\tau_{t,m}, k} 
& \mbox{o.w.}
\end{array} \right.
\label{eq:decoderEplus}
\end{equation}
\noindent where $\alpha_k$'s are the weights for the summation in 
(\ref{eq:decoderEplus}) based on the amount of pixel overlaps between
the designated predictor block and the error propagating MBs, and 
$\gamma < 1$ is the attenuation factor that reflects the dissipating 
effect of error in an earlier frame over a sequence of 
motion-compensated frames. In our simulations, $\gamma$ is 
chosen to be $0.9$.

If MB $m$ of frame $t$ is not correctly received, then we assume
the decoder performs a block copy from the same MB $m$ from
previous frame $t-1$. The resulting error is then error $e_{t-1,m}$
of MB $m$ in frame $t-1$, plus $\delta$, which is the change
in intensity between MBs $m$ in frame $t-1$ and $t$:
\begin{equation}
e^-_{t,m} = e_{t-1,m} + \delta
\label{eq:decoderEminus}
\end{equation}

$\delta$ can be estimated at decoder as follows. If there is a
corresponding block in texture map of the other captured viewpoint that 
is more reliably received, then this block is projected onto the current view
and $\delta$ is approximated as the difference between the projected block
and the co-located block within the previous frame of the current view. 
If there is no corresponding block in the other view or the corresponding block
is also erred, then we estimate $\delta$ as the difference among co-located
blocks of the previous two frames within the current view. For the initial 
frames in which one or two previous frames are not available, $\delta$ is 
estimated from spatially neighboring blocks with similar disparity values 
(i.e., of similar depth and hence likely the same physical object).

\subsection{Estimating Disparity Error at Decoder}

Similarly, we can compute the error in left disparity pixel
$Y^0_t(i,j^0)$ (or right disparity pixel $Y^1_t(i,j^1)$) by estimating 
the error in the MB $m=(i,j^0)$ (or $m=(i,j^1)$ for right disparity
pixel) that contains it.
We can estimate the disparity error\footnote{If a depth block
is predicted from the corresponding texture block of the same view
as done in \cite{merkle12}, (\ref{eq:decoderEplus}) can be easily
modified accordingly to reflect the inter-component dependency.}
 $\epsilon_{t,m}$ in MB $m$ 
using recursion as done for $e_{t,m}$ in (\ref{eq:decoderE}). In this case,
derivation of disparity errors due to correctly received and lost MBs are
analogous to the presented in (\ref{eq:decoderEplus}) and
(\ref{eq:decoderEminus}), respectively, with the exception that $\delta$ is
always estimated from co-located blocks within the previous two
frames.  %reply R3 C2

The effect of a disparity error on the synthesized pixel $(i,j)$, however,
is more indirect; a disparity error causes the wrong corresponding 
pixel $X^0_t(i,j^\#)$ to be used for synthesis of pixel $(i,j)$ in 
virtual view $S^v_t$. Now if the pixel patch around the local neighborhood
of texture pixel $X^0_t(i,j^0)$ is monotone, then using a texture pixel 
from a slightly off location $(i,j^{\#})$ will result in very small
increase in synthesized distortion. On the other hand, if the patch 
around texture pixel $(i,j^o)$ has high texture variation, then the 
resulting synthesized distortion can be significant.

\subsection{Proposed View Synthesis}

Having computed the estimated texture error $e_{t,m}$ and 
disparity error $\epsilon_{t,m}$ for the two MBs $m$ that contain 
corresponding texture pixel $X^0_t(i,j^0)$ and depth pixel $Y^0_t(i,j^0)$ 
of the left view respectively, as described in the past two sections, we 
now derive the \textit{worst case distortion} $d^0(i,j^0)$ for left 
corresponding pixel $X^0_t(i,j^0)$. Then, left reliability term $r^0$ can 
be subsequently computed. Together with right reliability term $r^1$, 
they determine how the two corresponding texture pixels $X^0_t(i,j^0)$ 
and $X^1_t(i,j^1)$ should be reweighed for our adaptive, error-aware
pixel blending.

Because disparity error leads to the selection of a wrong texture
pixel for blending, given the estimated disparity error $\epsilon_{t.m}$, 
we will consider the possible adverse effect of using texture pixel 
$X^0_t(i,l)$ for synthesis instead of $X^0_t(i,j^0)$, where $l$ in 
the range $[j^0 - \epsilon_{t.m} ,j^0 + \epsilon_{t.m}]$. Specifically,
for each texture pixel $X^0_t(i,l)$, we consider both the pixel-to-pixel 
texture intensity difference $|X^0_t(i,l)-X^0_t(i,j^0)|$ 
and the estimated texture error due to channel losses $e_{t,b(i,l)}$ 
for texture pixel $X^0_t(i,l)$. Mathematically, the worst-case 
distortion $d^0(i,j^0)$ for texture pixel $X^0_t(i,j^0)$ considering all 
pixels $(i,l)$ in the range is computed as follows:

\small
\begin{equation}
d^0(i,j^0) = \max_{l = j^0 - \epsilon_{t,m}, \ldots, j^0 + \epsilon_{t,m}} 
\left\{ e_{t,b(i,l)} + |X^0_t(i,l) - X^0_t(i,j^0)|  \right\}
\label{eq:worstErr}
\end{equation}
\normalsize

As done for $\delta$, the pixel intensity difference
$|X^0_t(i,l) - X^0_t(i,j^0)|$ in (\ref{eq:worstErr}) can be estimated 
at decoder using either corresponding pixels in texture map of the other 
captured view that is more reliably received, or MBs of previous 
correctly received frames of the same view with similar disparity values.

Having derived worst case distortion $d^0(i,j^0)$ and $d^1(i,j^0)$ for 
texture pixels, $X^0_t(i,j^0)$ and $X^1_t(i,j^1)$, we now define 
a reliability metric, $r^0$ and $r^1$, inversely proportional to distortion for the two pixels as follows:

\begin{equation}
r^0 = w \left( \frac{\kappa}{d^0(i,j^0) + \kappa} \right).
\end{equation}

\noindent In our implementation, $\kappa = d^1(i,j^0) + c$ such that distorion from both views are taken into consideration. $c$ is a small positive constant chosen so that $r^0$ is well defined even if both distortions are zero and $w$ is a scaling factor so that $r^0 + r^1 = 1$.

\section{Block-level Reference Picture Selection}
\label{sec:formulate}

Having described the error concealment strategy deployed during
view synthesis at the receiver, we now turn our attention to
optimization at the sender. 
At the sender side, during real-time encoding of both texture and 
depth maps, the encoder has the flexibility to select any  
MB in any past coded frame for motion compensation (MC) to encode
each MB in a current frame $t$. Using a well matched MB from the 
immediate previous frame $t-1$ for MC would lead to small encoding
rate, but may result in large expected distortion at receiver
due to a possibly long dependency chain of differentially coded MBs.
Using a MB from a frame further into the past for MC (or an intra-coded
MB), will lead to a larger encoding rate, but will also result
in a smaller expected distortion. Exploiting this flexibility of 
\textit{reference picture selection} (RPS), the goal is to pro-actively
minimize the overall expected distortion of an intermediate view at 
instant $t$, synthesized via DIBR using texture and depth maps of two 
adjacent coded views at decoder as described in the previous section, 
and subject to a transmission rate constraint. Leveraging our 
derivation for estimated texture and disparity errors at the receiver
in the previous section, we first discuss how synthesized distortion 
in an interpolated view is estimated % in a computation-efficient way 
at the sender. 
We then present the mathematical formulation of our proactive-feedback
block-level RPS optimization.

\subsection{Estimating Texture \& Disparity Error at Sender}

We first estimate the expected texture error $e_{t,m}$ of a MB $m$ 
in frame $t$ at sender in differentially coded texture video as follows.
Let $e_{t,m}(\tau_{t,m}, v_{t,m})$ be the texture error given it is 
motion-compensated using a block identified by MV $v_{t,m}$ inside 
a previous transmitted frame $X_{\tau_{t,m}}$, $\tau_{t,m} < t$. 
Let $p$ be the probability that MB $m$ is correctly 
\textit{received}. Similar to (\ref{eq:decoderE}), we can write
$e_{t,m}(\tau_{t,m}, v_{t,m})$ in terms of $e^+_{t,m}(\tau_{t,m} ,v_{t,m})$ 
and $e^-_{t,m}$, the expected texture error of MB $m$ if it is correctly 
received and lost, respectively. Unlike (\ref{eq:decoderE}), 
the expression is now probabilistic and depends on $p$, because
the delivery status of packet that contains MB $m$ is not known
at sender before it is transmitted:
\begin{equation}
e_{t,m}(\tau_{t,m}, v_{t,m}) = p ~ e^+_{t,m}(\tau_{t,m}, v_{t,m}) + 
(1-p) ~ e^-_{t,m}
\label{eq:encoderE}
\end{equation}
$e^+_{t,m}(\tau_{t,m} ,v_{t,m})$ and $e^-_{t,m}$ can be computed
recursively using previously derived (\ref{eq:decoderEplus}) and
(\ref{eq:decoderEminus}), respectively. In this case, $\delta$ is the difference
between the current block and the co-located in the previous frame.
Expected disparity error $\epsilon_{t,m}(\rho_{t,m}, u_{t.m})$ given
MV $u_{t,m}$ in previous disparity frame $Y_{\rho_{t,m}}$ can be computed
in a similar fashion.

\subsection{Computing Expected Synthesized View Distortion at Encoder}
\label{subsec:encoderDistortion}

Having derived expected texture and disparity errors $e^0_{t,m}$ and 
$\epsilon^0_{t.m}$ for a given MB $m$ in texture and disparity frames 
$X^0_t$ and $Y^0_t$ of view $0$, we now analyze the adverse effects of 
these errors to expected distortion in the synthesized image $S^v_t$ 
of virtual view $v$ at receiver.
One way to compute the expected synthesized distortion is 
to use a similar distortion expression as one derived in 
(\ref{eq:worstErr}) for receiver that ties both error terms together in 
a non-trivial way. However, this will introduce inter-dependency
among MBs in texture and depth maps, which will render the subsequent
block-level RPS optimization very difficult. 

\begin{figure*}%[htb]

\begin{minipage}[b]{.32\linewidth}
 \includegraphics[width=5.5cm,height=4.2cm]{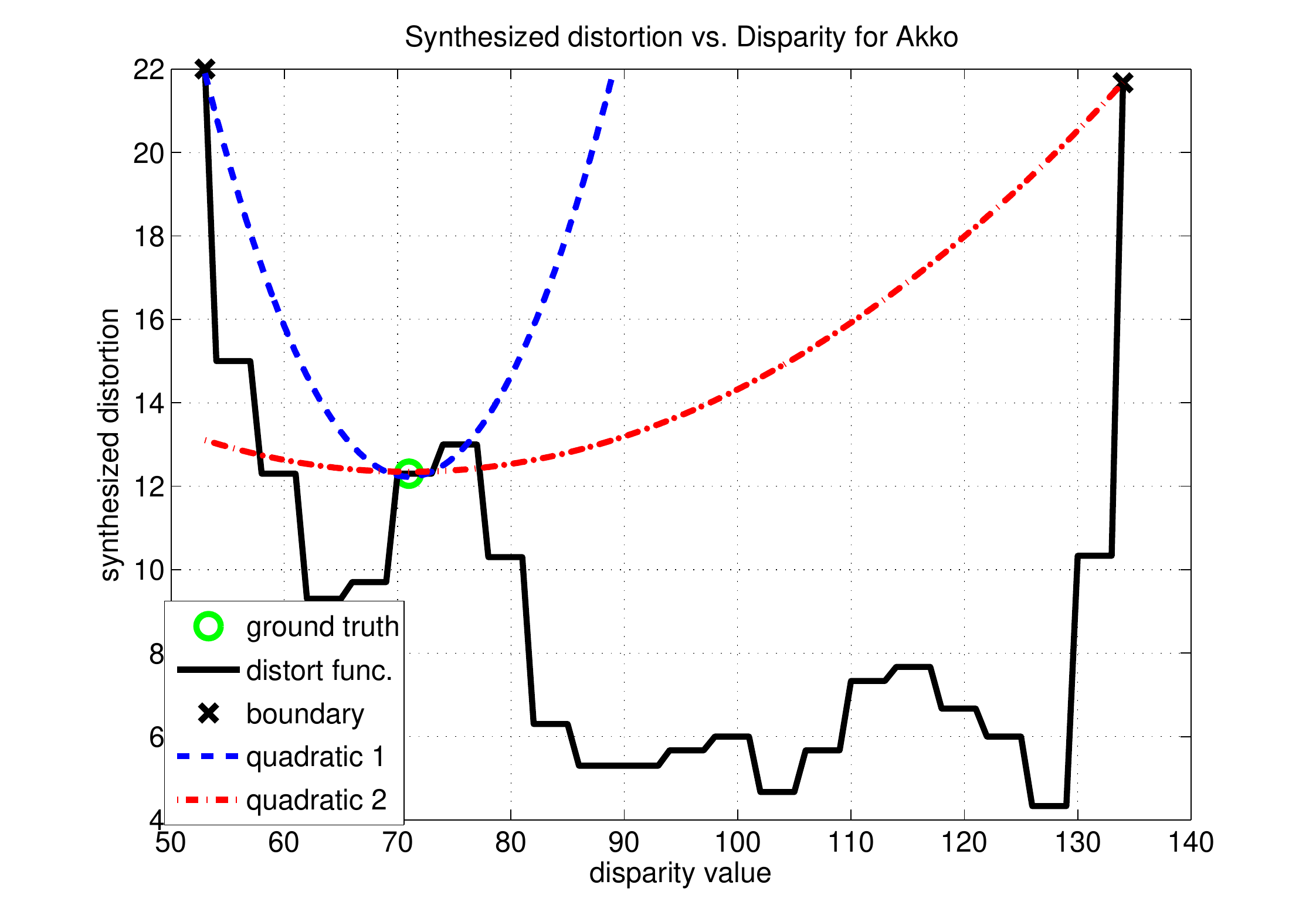}
 %\centerline{\epsfig{figure=akko_penalty.eps,width=5.5cm,height=4.2cm}}
  \centerline{\small{(a) Synthesized distortion functions}}\medskip
\end{minipage}
%\hfill
\begin{minipage}[b]{0.32\linewidth}
  \includegraphics[width=4.3cm]{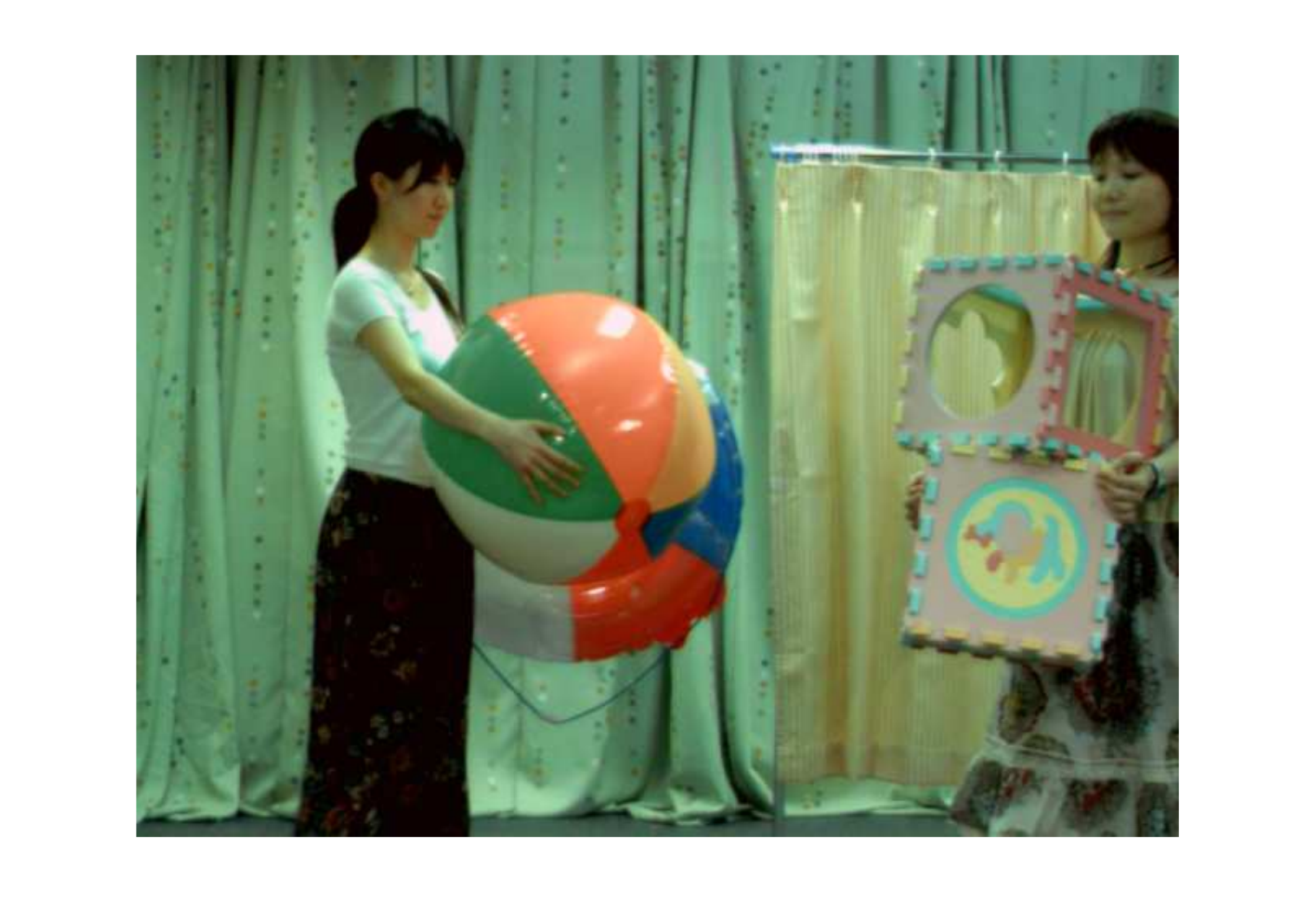}
  %\centerline{\epsfig{figure=akko_color.eps,width=4.3cm}}
  \vskip 1.2em
  \centerline{\small{(b) original frame}}\medskip
\end{minipage}
%\hfill
\begin{minipage}[b]{0.32\linewidth}
    \includegraphics[width=5.8cm]{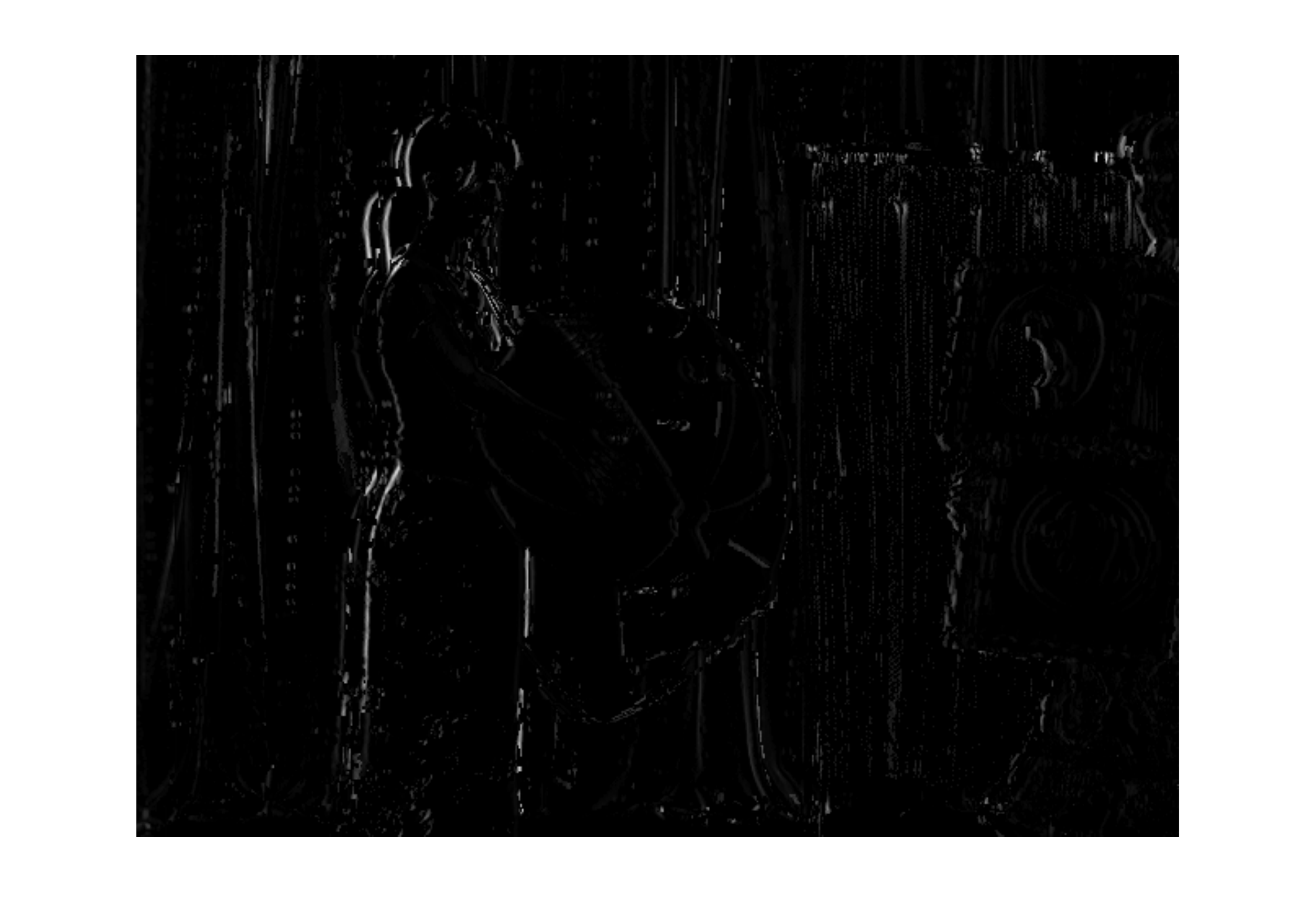}
 % \centerline{\epsfig{figure=akko_curvature.eps,width=5.8cm}}
  \centerline{\small{(c) Per-pixel curvatures}}\medskip
\end{minipage}
\caption{Synthesized distortion and quadratic model functions for one
  pixel are shown in (a) for image \texttt{Akko and Kayo} view 47 in (b). The resulting curvatures 
for for entire image are shown in (c).}
\label{fig:penalty}
\end{figure*}

Instead, we will use the following simplification. Texture 
error $e^0_{t,m}$, as discussed in Section~\ref{subsec:estTexErr},
contributes directly to the synthesized distortion. 
To estimate the ill effects of disparity error $\epsilon^0_{t,m}$
to synthesized distortion, we model synthesized distortion
as a quadratic function $g^0_{t,m}(~)$ of disparity
error $\epsilon^0_{t,m}$:
\begin{equation}
g^0_{t,m}(\epsilon^0_{t,m}) = \frac{1}{2} a^0_{t,m} 
\left( \epsilon^0_{t,m} \right)^2
\label{eq:penalty}
\end{equation}
\noindent where $a^0_{t,m}$ is the single parameter that describes the
curvature of the quadratic function $g^0_{t,m}(\epsilon^0_{t,m})$. 

This quadratic modeling of synthesized distortion was first used
in \cite{cheung11icip} for depth map coding. The key idea in
the model is to capture the synthesized view distortion sensitivity 
to disparity error in MB $m$ in a single parameter $a^0_{t,m}$. 
Specifically, we perform the following procedure for each disparity 
pixel in the block. We first construct a true synthesized distortion 
function for each pixel describing the resulting distortion as disparity
value deviates from the ground truth value. Then, we identify the
nearest \textit{boundary disparity values} below and above the ground 
truth where the corresponding synthesized distortion exceed a pre-defined 
threshold. Using ground truth and each of boundary disparity values, one
can construct two quadratic functions, or parabolas, with slope zero at ground truth. The curvature
of the sharper of the two parabolas is then chosen, and parameter 
$a^0_{i,m}$ in (\ref{eq:penalty}) is the average of all chosen curvatures
of pixels in the block. See Fig.~\ref{fig:penalty} for an example
and \cite{cheung11icip} for more details.

In general,
if the textural area corresponding to the depth MB $m$ is smooth and
inside a physical object, then synthesized distortion will not be sensitive 
to disparity error. The reason is that small error $\epsilon^0_{t,m}$
leading to mapping of wrong texture pixels of similar intensities
in the local neighborhood will result in only small synthesized distortion.
On the other hand, if the textural area corresponding to the depth MB
$m$ has large frequency contents inside a physical object or contains
boundary pixels between foreground object and the background, 
then synthesized distortion will be sensitive to disparity error.

\subsubsection{Computing block curvature $a^0_{t,m}$}

Specifically, we compute parameter $a^0_{t,m}$ for each MB $m$
as follows.
Distortion $d^0_{t,m}$ of MB $m$ is computed as a sum of its constituent 
pixels $(i,j)$'s distortions. As discussed in 
Section~\ref{subsec:standardVS}, a texture pixel $X^0_t(i,j)$ in the 
left texture map can be mapped to a shifted pixel 
$X^1_t(i,j - Y^0_t(i,j)*\eta)$ in the right texture map. 
We thus express synthesized distortion $d^0_{t,m}(\epsilon^0_{t,m})$ of
MB $m$ as the sum of the differences in texture pixel values 
between left pixels $X^0_t(i,j)$'s and mapped right pixels 
$X^1_t(i, j - (Y^0_t(i,j)+ \epsilon^0_{t,m})*\eta)$'s
due to disparity error $\epsilon^0_{t,m}$:

\small
\begin{equation}
d^0_{t,m}(\epsilon^0_{t,m}) = \sum_{(i,j) \in MB_m} | X^0_t(i,j) - 
X^1_t(i, j - (Y^0_t(i,j)+ \epsilon^0_{t,m})*\eta) |
\end{equation}
\normalsize

The synthesized distortion function $d^0_{t,m}(\epsilon^0_{t,m})$ for a pixel of view 
47 of the multiview image sequence \texttt{Akko and Kayo}~\cite{NagoyaSS} is shown in black in 
Fig.~\ref{fig:penalty}(a). 
We see that as the disparity error $\epsilon^0_{t,m}$ increases, i.e., the depth value moves away from ground truth, the distortion also increases generally. 
Our model $g^0_{t,m}(\epsilon^0_{t,m})$ will simply be the sharpest of the two quadratic functions
fit to $d^0_{t,m}(\epsilon^0_{t,m})$ as illustrated in Fig.~\ref{fig:penalty}(a). As an example, in Fig.~\ref{fig:penalty}(b) and (c), the original 
frame from view 47 of \texttt{Akko and Kayo} and the per-pixel 
curvatures of the quadratic model functions of the corresponding depth map are 
shown. We can clearly see that larger curvatures (in white) 
occur at object boundaries, agreeing with our intuition that a synthesized 
view is more sensitive to depth pixels at object boundaries.

Combining synthesized
distortion due to disparity error $\epsilon^0_{t,m}$ with that of texture error $e^0_{t,m}$, we can write the block-level
RPS optimization for MB $m$ in texture and depth maps of view $0$ as:

\begin{eqnarray}
\min_{\tau^0_{t,m}, v^0_{t,m}, \rho^0_{t,m}, u^0_{t,m}}
 \bar{D}^0_{t,m}(\tau^0_{t,m}, v^0_{t,m}, \rho^0_{t,m}, u^0_{t,m}) = \nonumber\\
  e^0_{t,m}(\tau^0_{t,m}, v^0_{t,m}) + 
g^0_{t,m}(\epsilon^0(\rho^0_{t,m}, u^0_{t,m}))
\label{eq:senderD}
\end{eqnarray}

\noindent where $\bar{D}^0_{t,m}$ denotes the expected distortion for a MB in
frame $t$. In words, (\ref{eq:senderD}) says that the expected
distortion for MB $m$ is a simple sum of: i) expected texture
error $e^0_{t,m}$, and ii) a quadratic term of the expected
disparity error $\epsilon^0_{t,m}$. This simple linear model for
distortion will play a significant role in simplifying the 
to-be-discussed RPS optimization procedure.

\subsubsection{Consideration for Receiver's Error Concealment}

Because of the error concealment scheme performed at decoder for
MBs that are visible from both captured views, as discussed in the 
previous section, distortion due to errors in MB $m$ of 
frames $X^0_t$ and $Y^0_t$ in view $0$ will contribute to
the actual synthesized view distortion \textit{only if} 
errors in the corresponding pixels in frames $X^1_t$ and $Y^1_t$ of view 
$1$ are worse. Otherwise, pixels in $X^1_t$ and $Y^1_t$ will be reweighted 
more heavily in the view synthesis process. 
Given the above observation, we can rewrite the expected 
synthesized view distortion $D^0_{t,m}$ in (\ref{eq:senderD}) due to 
texture and depth errors in MB $m$ of texture and depth frames $t$ of 
view $0$, as:

\begin{align}
D^0_{t,m}&(\tau^0_{t,m}, v^0_{t,m}, \rho^0_{t,m}, u^0_{t,m}) = \nonumber\\
&\min \left\{ ~e^0_{t,m}(\tau^0_{t,m}, v^0_{t,m}) \right.
+ g^0_{t,m}(\epsilon^0_{t,m}(\rho^0_{t,m},u^0_{t,m})), \nonumber\\
&\max_{k \in S^1_{m}} \left[
e^1_{t-1,k} + g^1_{t-1,k}(\epsilon^1_{t-1,k}) + \delta \right] 
\left.\right\}
\label{eq:encoderD}
\end{align}

\noindent where $S^1_m$ is the set of MBs in view $1$ that contains pixels
corresponding to MB $m$ in view $0$. In words, (\ref{eq:encoderD})
states that the synthesized distortion due to errors in MB $m$ of
view $0$ is the smaller of distortion due to view 0
and distortion due to view 1. Note that we write distortion due to 
view 1 using errors in \textit{previous frames}
$X^1_{t-1}$ and $Y^1_{t-1}$ plus $\delta$, so that
$D^0_{t,m}$ will not have dependency on MVs of current frames
$X^1_t$ and $Y^1_t$ of view 1. (Recall $\delta$ is the change in
intensity between MBs in frame $t-1$ and $t$.)
This avoids inter-dependency
of free variables, simplifying the to-be-discussed
optimization algorithm.

\subsection{Block-level Reference Picture Optimization}

We are now ready to formally define our optimization for block-level
reference picture selection. Our objective is to minimize
the sum of induced synthesized distortion from MB $m$ in both
view $0$ and $1$:

\begin{align}
\min_{\{ \tau^i_{t,m}, v^i_{t,m}, \rho^i_{t,m}, u^i_{t,m} \}} 
\sum_{i \in \{0,1\}} \left(\right.
\sum_{m \in \bar{\mathcal{L}_t}} 
\bar{D}^i_{t,m}(\tau^i_{t,m}, v^i_{t,m}, \rho^i_{t,m}, u^i_{t,m}) \nonumber\\
+ \sum_{m \in \mathcal{L}_t} 
D^i_{t,m}(\tau^i_{t,m}, v^i_{t,m}, \rho^i_{t,m}, u^i_{t,m})
\left.\right)
\label{eq:obj}
\end{align}

\noindent where $\bar{\mathcal{L}}_t$ and $\mathcal{L}_t$ are the set of of
MBs in frame $t$ without and with corresponding MBs in the opposing
captured view, respectively.

We note that (\ref{eq:obj}) is an approximation of the actual
synthesized distortion at intermediate view $S^v_t$.
%since in general, texture error $e_{t,m}$ and disparity error 
%$\epsilon_{t,m}$ affect synthesized distortion in a complicated, 
%non-linear way, especially when both $e_{t,i}$ and $\epsilon_{t,i}$ are
%large. 
Nonetheless, (\ref{eq:obj}) is a good approximation when only one of 
the two errors $e^i_{t,m}$ and $\epsilon^i_{t,m}$ is non-zero, and 
it leads to a simple optimization procedure as discussed in 
Section~\ref{subsec:algo}.

The optimization is subject to the rate constraint $R_t$ at instant $t$:

\begin{equation}
\sum_{i \in \{0, 1\}} \sum_m 
r^i_{t,m}(\tau^i_{t,m}, v^i_{t,m}) + \zeta^i_{t,m}(\rho^i_{t,m}, u^i_{t,m}) 
\leq R_t
\label{eq:constr}
\end{equation}

\noindent where $r^i_{t,m}$ and $\zeta^i_{t,m}$ are the resulting bit overhead required
to code MB $m$ in texture and depth frame of view $i$, $X^i_t$ and $Y^i_t$, 
respectively, given selection of reference frame / MV pair, 
$(\tau^i_{t,m}, v^i_{t,m})$ and $(\rho^i_{t,m}, u^i_{t,m})$, respectively.

\subsection{Optimization Algorithm}
\label{subsec:algo}

Instead of solving the constrained optimization problem (\ref{eq:obj})
and (\ref{eq:constr}), we can solve the corresponding Lagrangian 
problem instead for given multiplier $\lambda > 0$:

\begin{align}
\min_{\{ \tau^i_{t,m}, v^i_{t,m}, \rho^i_{t,m}, u^i_{t,m} \}}& 
\sum_{i \in \{0,1\}} \left( \right.
\sum_{m \in \bar{\mathcal{L}_t}}
\bar{D}^i_{t,m}(\tau^i_{t,m}, v^i_{t,m}, \rho^i_{t,m}, u^i_{t,m}) \nonumber\\
&+ \sum_{m \in \mathcal{L}_t}
D^i_{t,m}(\tau^i_{t,m}, v^i_{t,m}, \rho^i_{t,m}, u^i_{t,m})
\left.\right)\nonumber \\
&+ \lambda \sum_{i \in \{0,1\}} \sum_m \left( \right. r^i_{t,m}(\tau^i_{t,m}, v^i_{t,m}) \nonumber\\
&+ \zeta^i_{t,m}(\rho^i_{t,m}, u^i_{t,m})\left.\right)
\label{eq:lagrange}
\end{align}

To solve (\ref{eq:lagrange}) optimally, it is clear that we can separately
optimize each pair of MBs $m$ in texture and disparity frame $X^i_t$ and 
$Y^i_t$. For MB $m$ that has no corresponding MB in opposing view, 
i.e., $m \in \bar{\mathcal{L}_t}$:

\begin{align}
\min_{ \tau^i_{t,m}, v^i_{t,m}, \rho^i_{t,m}, u^i_{t,m} } 
\bar{D}^i_{t,m}(\tau^i_{t,m}, v^i_{t,m}, \rho^i_{t,m}, u^i_{t,m})~~~~~~\nonumber\\ 
+ \lambda~ \left[~r^i_{t,m}(\tau^i_{t,m}, v^i_{t,m}) + 
\zeta^i_{t,m}(\rho^i_{t,m}, u^i_{t,m}) \right] ~~~ \forall i, m
\label{eq:lagrangeSep}
\end{align}

In this case, $\bar{D}^i_{t,m}$ separates into two terms, $e^i_{t,m}$
and $g^i_{t,m}$, and the texture and depth map MB variables 
can clearly be optimized separately.

For MB $m$ that has a corresponding MB in opposing view, 
$\bar{D}^i_{t,m}$ is replaced by $D^i_{t,m}$ in (\ref{eq:lagrangeSep}).
There is now an inter-dependency between texture and depth map MB 
variables in the distortion term.
We can simplify (\ref{eq:lagrangeSep}) into the following
two equations, where $(\tau^{i,*}_{t,m}, v^{i,*}_{t,m})$ means the 
best MV that minimizes texture error $e^i_{t,m}$ only, so
that the searches for MV for texture and depth maps can be 
performed separately:

\small
\begin{align}
\min_{ \tau^i_{t,m}, v^i_{t,m} }&D^i_{t,m}(\tau^i_{t,m}, v^i_{t,m}, \rho^{i,*}_{t,m}, u^{i,*}_{t,m}) 
+ \lambda~ r^i_{t,m}(\tau^i_{t,m}, v^i_{t,m}) 
 ~~~ \forall i, m \nonumber \\
\min_{ \rho^i_{t,m}, u^i_{t,m} }&
D^i_{t,m}(\tau^{i,*}_{t,m}, v^{i,*}_{t,m}, \rho^{i}_{t,m}, u^{i}_{t,m}) 
+ \lambda~ \zeta^i_{t,m}(\rho^i_{t,m}, u^i_{t,m})
 ~~~ \forall i, m\nonumber\\
\label{eq:lagrangeSep2}
\end{align}
\normalsize

Equation (\ref{eq:lagrangeSep2}) is an approximation to (\ref{eq:lagrangeSep})
in the following sense. If errors in the opposing views in
(\ref{eq:encoderD}) are relatively large, then errors $e^i_{t,m}$ and 
$\epsilon^i_{t,m}$ can indeed be separately optimally traded off with
their respective rate terms $r^i_{t,m}$ and $\zeta^i_{t,m}$, as done
in (\ref{eq:lagrangeSep2}), with no loss in optimality. If errors in
the opposing views are relatively small, then errors in current view 
can actually be ignored. Hence by setting MV of texture map to
be best error-minimizing one $(\tau^{i,*}_{t,m}, v^{i,*}_{t,m})$ when 
optimizing MV of disparity map, we are forcing the optimization to 
focus on minimizing induced synthesized distortion unless
errors in opposing views are very small comparatively. This 
conservative approach ensures the combined resulting induced
synthesized distortion from texture and disparity errors 
will not be large. (\ref{eq:lagrangeSep2}) is minimized by searching through all 
feasible MVs in all valid reference frames. This can be
done efficiently, for example, in a parallel implementation. 

\section{Complexity Analysis}
\label{sec:complexity}

The proposed strategies introduce additional complexity costs for both the
sender and receiver of the free-viewpoint conferencing system. At the sender
side, the proposal with largest computational load is the determination of the
curvature parameters representing sensitivity to disparity errors employed in
(\ref{eq:penalty}). Though in our current implementation we compute
$a^0_{t,m}$ for block $m$ in (\ref{eq:penalty}) by first painstakingly 
computing curvature parameters for each pixel in the block as described
in Section~\ref{subsec:encoderDistortion}, one can approximate $a^0_{t,m}$
in a computation-efficient manner by simply examining the high frequency
DCT components of block $m$ and its neighboring blocks in texture and 
depth maps (high frequencies in depth block indicate an edge, while high 
frequencies in texture block indicate fast changing textural content.) 
As fast curvature parameter approximation is not the main focus of this 
paper, we simply assume curvature parameters can be computed efficiently 
and discuss complexity of our proposed optimization procedures.

Still at sender, the loss-resilient coding proposals encompass the estimation
of texture/disparity errors and the optimization of RPS. Texture error
estimation involves determining $e^+$ and $e^-$ as combined in
(\ref{eq:encoderE}) and defined by (\ref{eq:decoderEplus}) and
(\ref{eq:decoderEminus}), respectively.  When correctly received, $e^+$ of a
non-intra MB of size $N \times N$ involves a weighted average over all the MB
pixels and thus $N \times N$ multiplication operations. The error $e^-$ of an
incorrectly received MB is a function of $\delta$ which is taken as the
difference between the current MB and the co-located MB of the previous
frame. This difference contributes in $N \times N$ subtraction operations to
the complexity count. Complexity analysis of disparity error estimation for
encoding provides similar results. Lastly, the optimization algorithm,
described in sub-section \ref{subsec:algo}, considers the tradeoff between an
expected distortion measure comprised of texture and disparity errors across
both views and rate terms for selecting reference pictures. It is used in
substitution of the standard rate-distortion optimization employed in coding
of texture and depth and thus imposes no additional computational complexity
costs. The aforementioned coding proposals represent a modest complexity
increment in terms of arithmetic operations. In our implementation, an average
increase of only $1.6\%$ in execution time with respect to coding with
reactive feedback channel was registered. Note that for these simulations fast
motion estimation was used and file I/O operations, used in our
implementation, were not considered.

At the receiver, texture and disparity error estimation for decoding present a
computational cost similar to the one determined for the sender. Nonetheless,
when decoding, the estimation of $\delta$ in (\ref{eq:decoderEminus}) is
dependent on whether packets of both views are lost simultaneously or whether
packets in only one view are lost. In the former case, delta is the difference
among co-located MBs in the two previous frames, representing $N \times N$
subtraction operations. In the latter case, prior to estimating $\delta$ as a
difference of MBs, the correctly received pixels from the adjacent view are
projected onto the view with losses, representing an extra $N \times N$
addition operations. Furthermore, at the receiver, the worst case distortion
defined in (\ref{eq:worstErr}) must be calculated for the virtual view prior
to adaptive blending. This calculation represents $N \times N \times
(2 \epsilon_{t,m}+1)$ subtraction operations. Here, the disparity error
$\epsilon_{t,m}$ is content dependent and generally small except along
boundary pixels with large intensity differences. The adaptive blending step,
based on worst case distortion estimates, is equivalent in complexity to the
standard blending procedure which it substitutes. In conclusion, our decoding
proposals also impose a modest complexity increment in terms of additional
arithmetic operations.

\section{Experimentation}
\label{sec:results}

We verify the contributions of our proposed error-concealment strategy during
view synthesis (adaptive blending) as well as our proposed block-level RPS 
encoding strategy (proactive-feedback) through extensive experiments in this section. We first
describe the experimental setup in which our proposals are compared, and 
then present streaming results in terms of an objective measure (PSNR) 
and subjective gains.

\subsection{Experimental Setup}

Our proposed view synthesis strategy, as described in
Section~\ref{sec:conceal}, is based on an error-aware \textit{adaptive
blending} procedure. Adaptive blending is implemented in the MPEG View
Synthesis Reference Software (VSRS v3.5)~\cite{Tanimoto08} and comparisons 
are drawn against the standard version of the software. 

In addition to evaluating the contributions of adaptive blending, the 
proposed proactive-feedback RPS encoding strategy is compared to two other 
RPS-based error-resilient alternatives. Note that in all RPS strategies 
we assume a feedback channel is present. The channel can transmit 
acknowledgment to the encoder, after a round-trip-time delay (RTT), of 
whether a packet has been correctly received or lost. In our experiments, 
RTT is set to be constant and fixed to $133$ ms ($4$ frame of delay at 
$30$ fps). 
 
The first RPS alternative reacts to feedback information by avoiding the 
use of loss-affected regions in past frames as reference for coding
of future MBs. This alternative is termed \textit{reactive feedback} and
serves as a baseline for comparisons. The second RPS alternative employed in
our tests is based on our previous work~\cite{bruno12-2}. Similar to our current
proposal, \cite{bruno12-2} consists of a proactive RPS scheme combined with
feedback
information, which estimates the reconstructed error of a block from given past frames, protecting those of
greater importance to synthesis. However, unlike our current proposal,
in~\cite{bruno12-2} views are encoded independently, \textit{i.e.}, expected
distortions of one view do not affect the encoding of the adjacent
view. Note that the experimental results reported in~\cite{bruno12-2} used a
different feedback channel which only acknowledged whether an entire frame was
received error-free. 

%In fact, for such proactive RPS strategies a feedback
%channel is not a requirement. \red{Dan: this is confusing to me: why won't we need a feedback in this case?  
%We need to be more clear what is the difference between the proactive RPS in [36] and our current proposal.} 
%Nevertheless, we use the feedback information in order
%to enhance the performance by confirming certain block errors and thus improving
%accuracy of our recursive distortion model.

From the described view synthesis and RPS encoding strategies, four setups are
tested in the following sections: 
\begin{itemize}
	\item standard blending with reactive feedback channel (RFC),
	\item standard blending with proactive RPS from~\cite{bruno12-2} with feedback information (RPS1),
	\item standard blending with the proposed proactive RPS with feedback information (RPS2) and
	\item the proposed adaptive blending with the proposed proactive RPS with feedback information (ARPS).
\end{itemize}

All results are evaluated in terms of synthesized view quality. 
Specifically, for each multi-view sequence, we choose three neighboring views, and transmit only the left
and right views, reserving the withheld center view as ground truth.  
We use test sequences from the Nagoya University database \cite{NagoyaSS}. The selected sequences are 
the first $150$ frames of: \texttt{Pantomime} ($1280\times960$ pixels), \texttt{Kendo} ($1024\times768$ pixels) and \texttt{Akko and Kayo}
($640\times480$ pixels), all at $30$ fps. For  \texttt{Pantomime}
view $40$ was synthesized using views $39$ and $41$, for  \texttt{Kendo} 
view $2$ is interpolated from view $1$ and $3$, and for  \texttt{Akko and Kayo} 
views $47$ an $49$ are used to synthesize view $48$. The view synthesis software~\cite{Tanimoto08}
was set to work with integer pixel precision. 

Currently, our encoding scheme, including error estimation, is implemented
only for $P16\times16$ mode in H.264/AVC JM reference software v18.0
\cite{JM180}. Therefore, the only modes available in all simulations are
$P16\times16$, Skip or Intra blocks (no block or sub-block partitions are
allowed).  More extensive comparison using larger number of available modes is
a subject of future study. Note that the appropriate $\lambda$ for
(\ref{eq:lagrangeSep2}) in all simulations was selected empirically. The QP
was set to $28$ in all setups. 

Four transport packets are used for each depth map frame, while twelve are
used for each texture frame packets due to higher associated bit
rates. Simulations include losses of $2\%$, $5\%$ and $8\%$ of the packets for
both texture and depth maps. In order to provide meaningful comparisons, the same packets are
lost in all schemes. 
We assume the picture parameter set (PPS) and
sequence parameter set (SPS) are reliably transmitted out-of-band.
%Loss can be avoided by retransmission or use of a more secure channel for
%such packets. 
Both depth maps and texture are encoded using $64$ pixel search window, CABAC entropy encoder
and $IPPP...$ encoding mode. When a MB is lost during transmission, the
co-located block from the previous frame is used in its place.

\subsection{Objective Streaming Performance}

In this subsection we present the results of our simulations using PSNR as an objective
quality metric. The bitrates for each sequence, for 
each percentage of lost packets, are presented in Table \ref{res:tab1}. The
bitrates are computed considering both views and depth maps. Note that, as defined by 
(\ref{eq:encoderE}), the probability of MB losses is considered during error estimation. 
Therefore, as this probability increases, the encoder raises the level
of protection in each MB resulting in higher bitrates as seen in Table \ref{res:tab1}. %reply R1 C8
The most suitable $\lambda$ in (\ref{eq:lagrangeSep2}) can be found via binary search where the optimization problem is solved multiple times (each time with a different $\lambda$) for each frame. In practice, the $\lambda$ that is used in previous frame can be reused, with an optional local adjustment if the resulting rate is too far below or above a rate constraint. In our experiments, we first run the RFC setup. Then, we select among various tested candidates the $\lambda$ for RPS1 and RPS2 which results in bitrate closest to that of RFC. Our goal was to match the bitrates between all setups in order to make a fair comparison between them.

\begin{table}
\small
\begin{center}
\caption{Bitrates used during Simulations (including both views and depth maps)}
\begin{tabular}{|l|l|l|}\hline
\textbf{Sequence}& \textbf{Loss Rate} & \textbf{Bitrate}\\\hline
\multirow{3}{*}{ \texttt{\footnotesize{Pantomime}}} & 2\% & 9.7 Mbps \\
 & 5\% &  10.2 Mbps  \\
 & 8\% &  11.2 Mbps  \\\hline
\multirow{3}{*}{ \texttt{\footnotesize{Kendo}}} & 2\% & 6.7 Mbps \\
 & 5\% & 6.8 Mbps \\
 & 8\% & 7.0 Mbps \\\hline
\multirow{3}{*}{ \texttt{\footnotesize{Akko and Kayo}}} & 2\% & 6.6 Mbps  \\
 & 5\% & 7.0 Mbps \\
 & 8\% & 7.5 Mbps \\\hline
\end{tabular}
\label{res:tab1}
\end{center}
\normalsize
\end{table}

A summary of the PSNR results obtained for synthesized views (assuming original views as ground truth) can be found in
Tables \ref{res:tab2} and \ref{res:tab3}. Table \ref{res:tab2} presents
the average PSNR over all frames for each sequence and error rate for all four tested setups: $RFC$, $RPS1$, $RPS2$ and
$ARPS$. Table \ref{res:tab3} shows the maximum PSNR gain for a single
frame in each simulation when comparing setups RPS1, RPS2 and ARPS to the use
of RFC.

We first observe that results RPS1, based on previous work \cite{bruno12-2}, already outperform the
use of RFC. In terms of average PSNR, RPS1 outperforms
RFC by as much as $0.48$ dB for
\texttt{Kendo} at $8\%$ error rate as shown in Table \ref{res:tab2}. For \texttt{Pantomime} and \texttt{Akko and Kayo} 
the largest average gains are of $0.42$ dB and $0.21$ dB,
respectively. Note that when comparing $RPS1$ to RFC, the gains increase as the
error rates increase for all sequences. For a single frame, shown in Table~\ref{res:tab3}, RPS1 can
outperform RFC by as much as $2.01$ dB 

The results obtained by RPS2 show an improvement over previous
work RPS1. However, improvements are modest since both schemes
use similar optimization processes. The main difference is that the
proposed encoding strategy in RPS2 protects each MB simultaneously in both views,
therefore the same level of protection is achieved with a smaller
bitrate. By combining our proposed encoding with our
proposed adaptive blending (ARPS) significant improvements are achieved,
specially for \texttt{Pantomime} and \texttt{Kendo}. From Table \ref{res:tab2}, we
observe that ARPS can outperform the use of a
RFC by as much as $0.82$ dB for \texttt{Kendo} at $8\%$
error rate. The maximum gain of ARPS with respect to 
RFC for a single frame, presented in Table \ref{res:tab3}, is $3.29$ dB for \texttt{Kendo}.

\begin{table}
\footnotesize
\begin{center}
%\caption{Average PSNR: $RFC$- Regular blending $+$ Reactive feedback channel, $RPS1$- Regular Blending $+$ RPS \cite{bruno12-2}, $RPS2$- Regular Blending $+$ new RPS encoding mode and $ARPS$- Adaptive blending $+$ new RPS encoding mode}
\caption{Average PSNR over all frames for tested setups.}
\begin{tabular}{|l|l|l|l|l|l|}\hline
\multirow{2}{*}{\textbf{Sequence}}& {\textbf{Loss}}& \multicolumn{4}{|c|}{\textbf{Avg. PSNR}}\\
 & {\textbf{Rate}} & \textbf{RFC} & \textbf{RPS1} & \textbf{RPS2} & \textbf{ARPS} \\\hline
\multirow{3}{*}{ \texttt{Pantomime}} & 2\% & 31.44 dB & 31.65 dB & 31.70 dB & 31.81 dB \\
 & 5\% & 30.02 dB & 30.33 dB & 30.44 dB & 30.62 dB\\
 & 8\% & 28.99 dB & 29.41 dB & 29.47 dB & 29.68 dB\\\hline
\multirow{3}{*}{ \texttt{Kendo}} & 2\% & 33.41 dB &  33.59 dB & 33.62 dB & 33.74 dB\\
 & 5\% & 31.79 dB &  32.23 dB &  32.32 dB & 32.45 dB\\
 & 8\% & 30.37 dB & 30.85 dB & 31.00 dB & 31.19 dB\\\hline
 \texttt{Akko} & 2\% & 28.08 dB & 28.15 dB & 28.19 dB & 28.22 dB\\
 \texttt{and}& 5\% & 26.85 dB & 26.97 dB & 27.03 dB & 27.06 dB\\
 \texttt{Kayo}& 8\% & 26.16 dB & 26.37 dB & 26.44 dB & 26.53 dB\\\hline
\end{tabular}
\label{res:tab2}
\end{center}
\normalsize
\end{table}

\begin{table}
\small
\begin{center}
%\caption{Maximum PSNR gain with respect to reactive feedback: $RPS1$- Regular Blending $+$ RPS \cite{bruno12-2}, $RPS2$- Regular Blending $+$ new RPS encoding mode and $ARPS$- Adaptive blending $+$ new RPS encoding mode}
\caption{Maximum PSNR gain for a single frame with respect to RFC.}
\begin{tabular}{|l|l|l|l|l|}\hline
\multirow{2}{*}{\textbf{Sequence}}& {\textbf{Loss}}& \multicolumn{3}{|c|}{\textbf{Max. PSNR Gain}}\\
 & {\textbf{Rate}} & \textbf{RPS1} & \textbf{RPS2} & \textbf{ARPS} \\\hline
\multirow{3}{*}{ \texttt{\footnotesize{Pantomime}}} & 2\% & 1.26 dB & 1.52 dB & 2.00 dB\\
 & 5\% & 0.86 dB & 1.01 dB & 1.45 dB\\
 & 8\% & 1.13 dB & 1.25 dB & 1.64 dB\\\hline
\multirow{3}{*}{ \texttt{\footnotesize{Kendo}}} & 2\% & 1.42 dB & 1.48 dB & 1.72 dB\\
 & 5\% & 1.69 dB & 1.81 dB & 2.94 dB\\
 & 8\% & 2.01 dB & 2.03 dB & 3.29 dB\\\hline
\multirow{3}{*}{ \texttt{\footnotesize{Akko and Kayo}}} & 2\% & 0.27 dB & 0.27 dB & 0.31 dB \\
 & 5\% & 0.50 dB & 0.55 dB & 0.64 dB\\
 & 8\% & 0.47 dB & 0.51 dB & 0.67 dB\\\hline
\end{tabular}
\label{res:tab3}
\end{center}
\normalsize
\end{table}

\begin{figure}[!htbp]
\centering
\subfigure[][]{\includegraphics[width=0.49\textwidth]{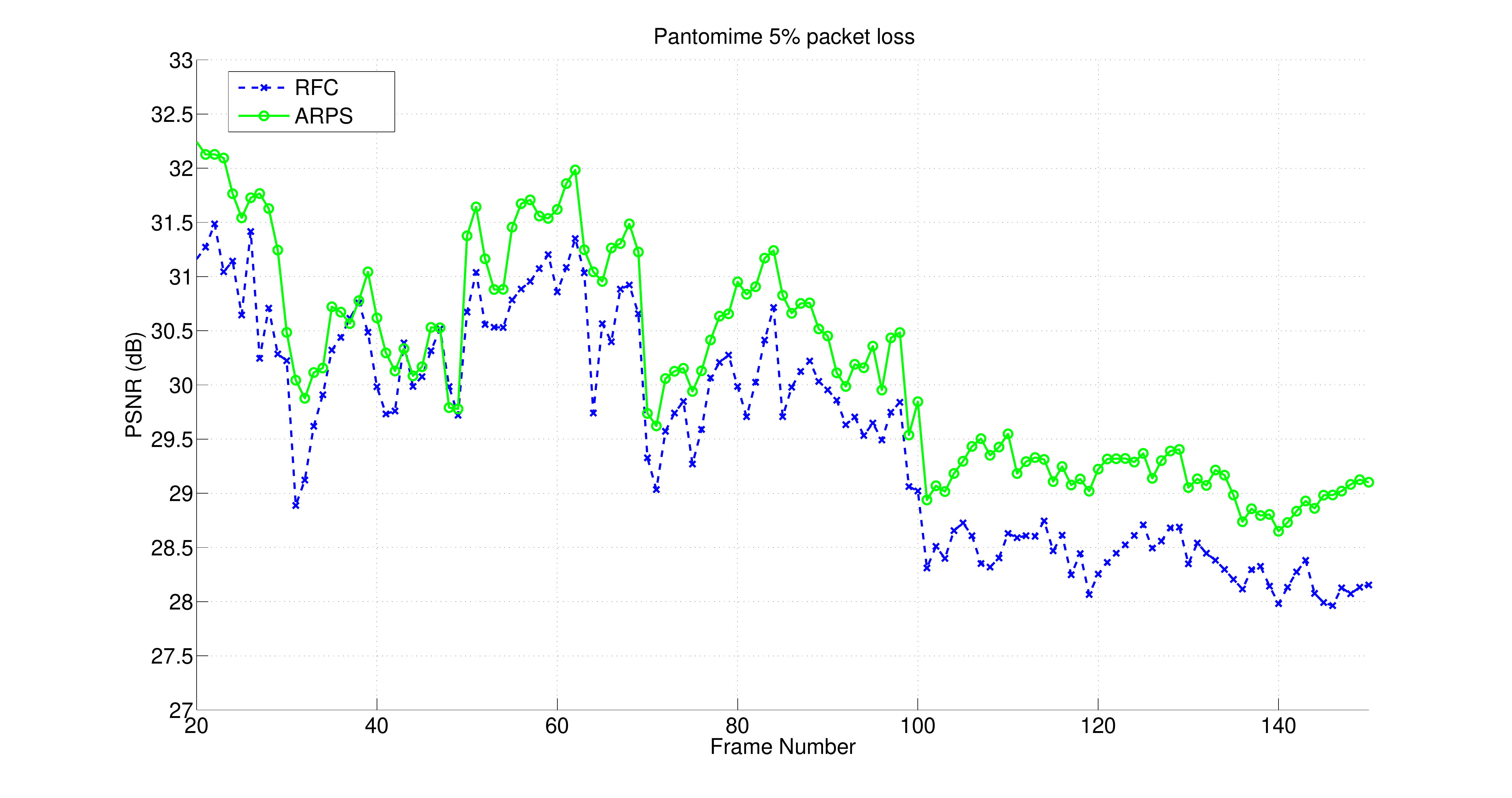}}\\
\subfigure[][]{\includegraphics[width=0.49\textwidth]{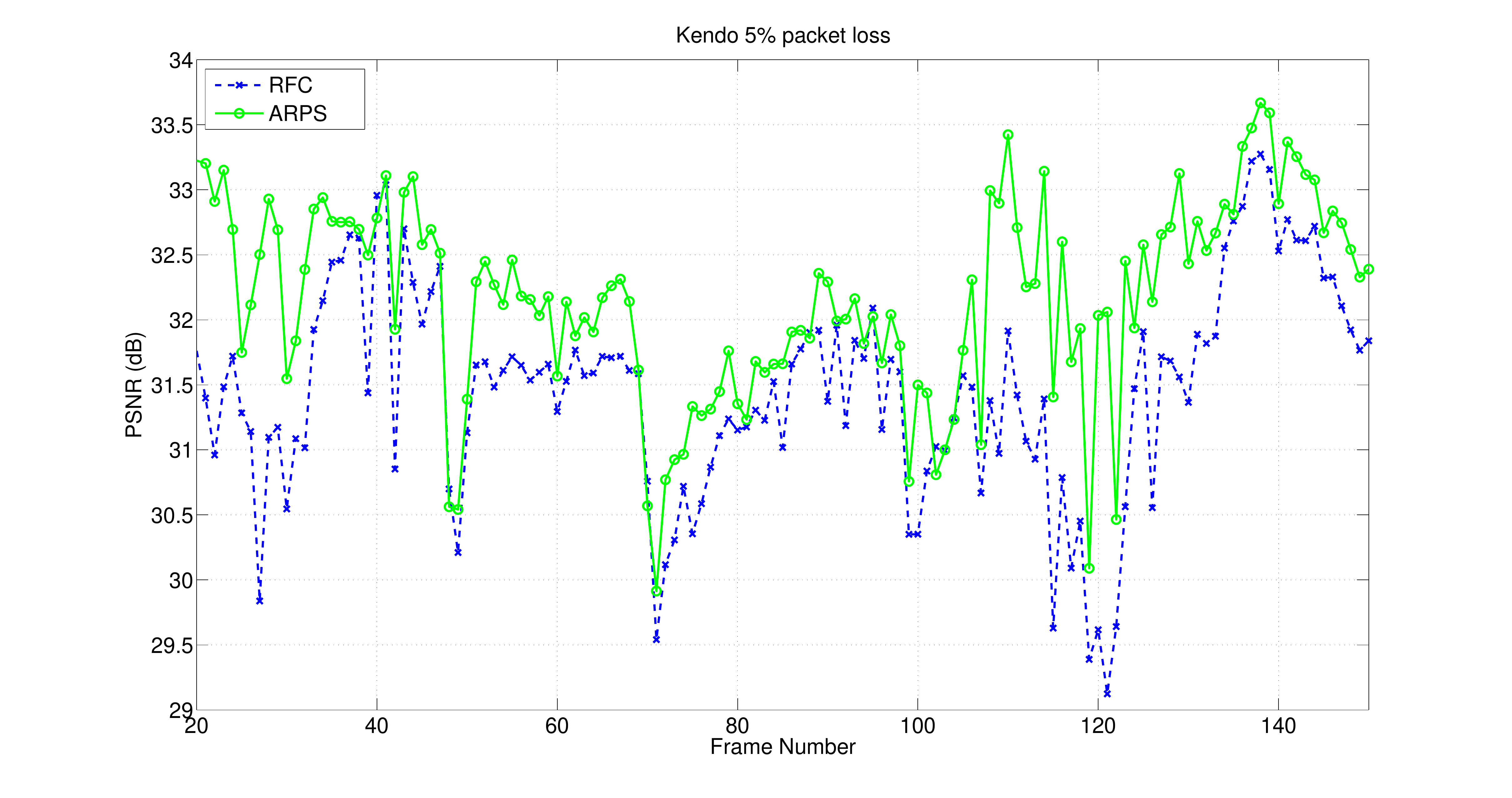}}\\
\subfigure[][]{\includegraphics[width=0.49\textwidth]{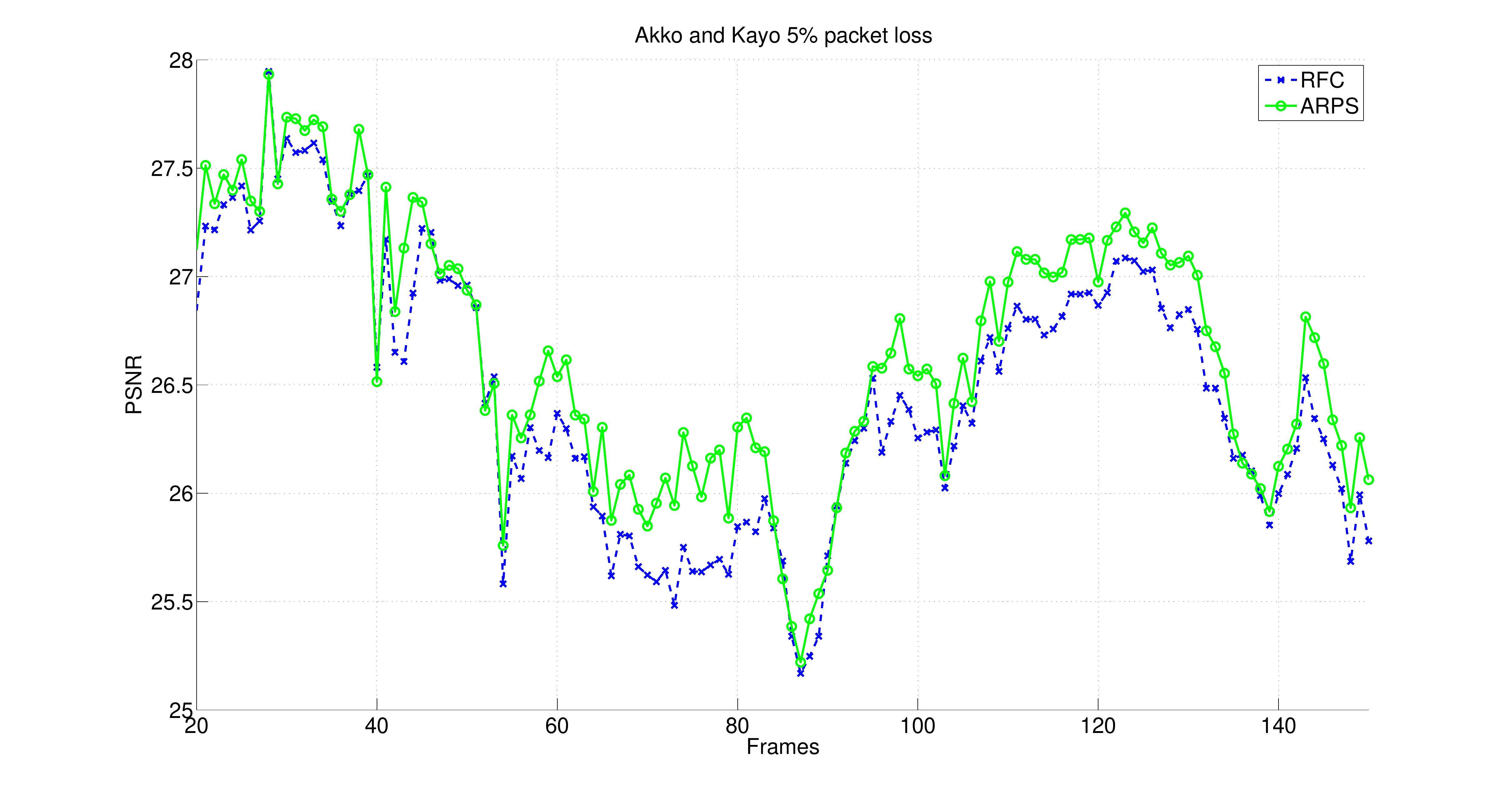} }
\vspace{-2mm}
\caption{PSNR per frame for ARPS and RFC at $5\%$ packet loss rate for (a) \texttt{Pantomime}, (b) \texttt{Kendo} and (c) \texttt{Akko and Kayo}.}
\label{fig:res}
\end{figure}

In Fig.~\ref{fig:res}, PSNR values for the proposed ARPS and the baseline RFC are presented for each frame. Both schemes consider a $5\%$ packet loss
rate and PSNR values are shown from frame 20 though frame 150 for better visualization. Due to the use of feedback, we see
that both schemes are generally able to avoid continuous propagation of errors
due to losses. Nevertheless, we see that our method can better
withstand the transient effect of packet losses by providing stronger
protection to more important regions. Furthermore, our ARPS consistently performs
above RFC. 

\subsection{Visual Gains}

\begin{figure}[!htbp]
\centering
\subfigure[][]{\includegraphics[width=0.49\textwidth]{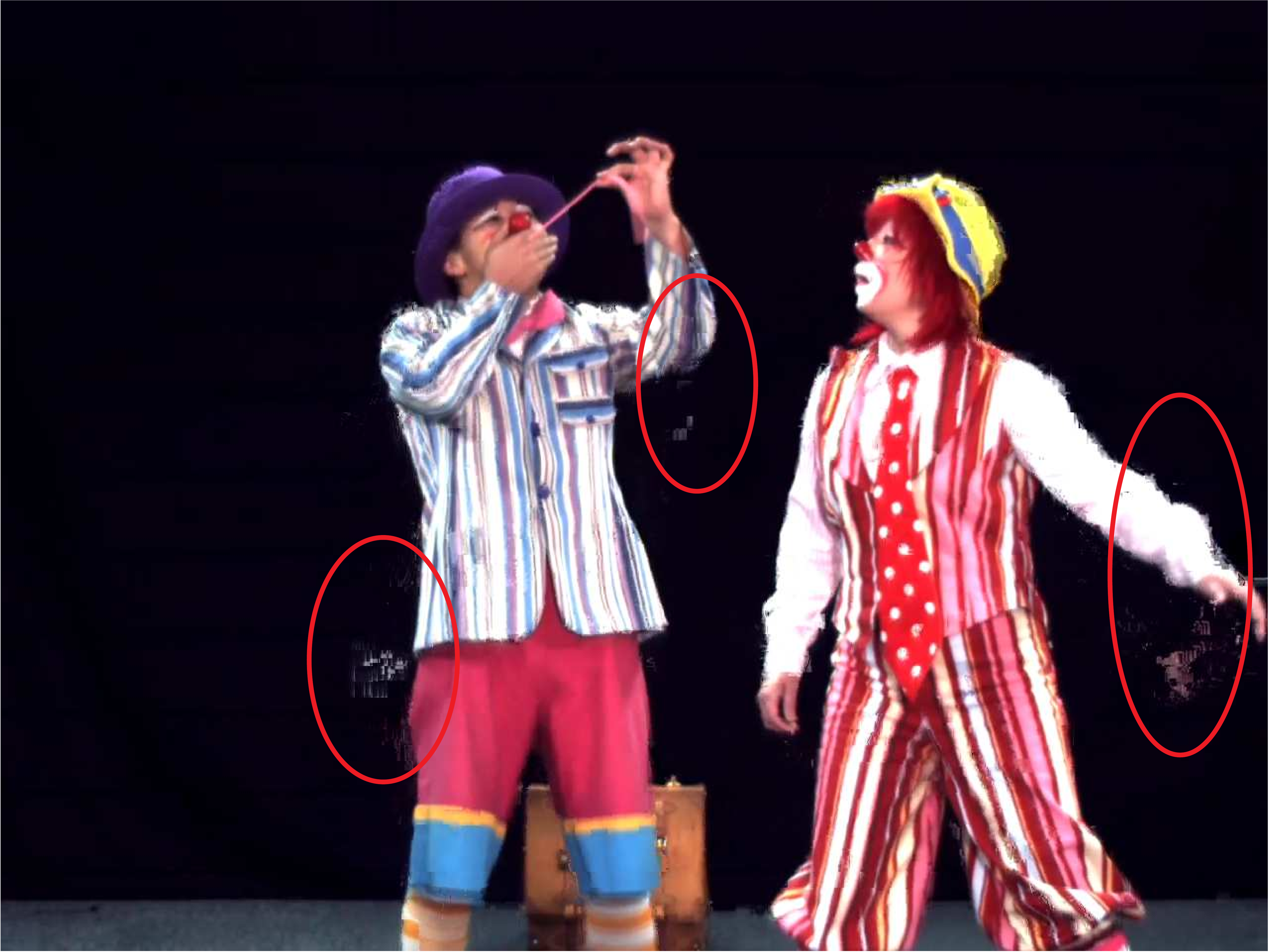}}\\
\subfigure[][]{\includegraphics[width=0.49\textwidth]{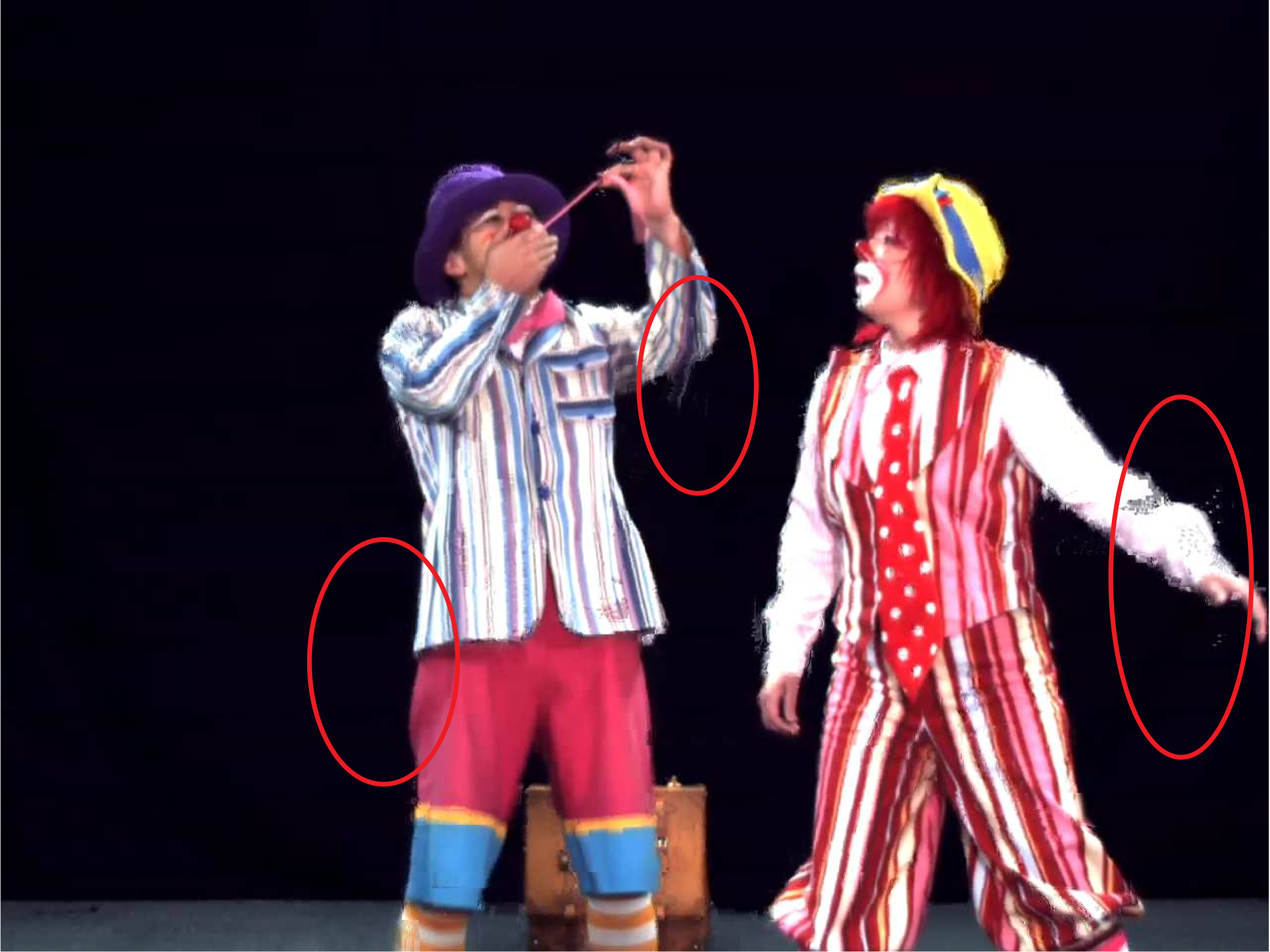}}\\
\subfigure[][]{\includegraphics[width=0.49\textwidth]{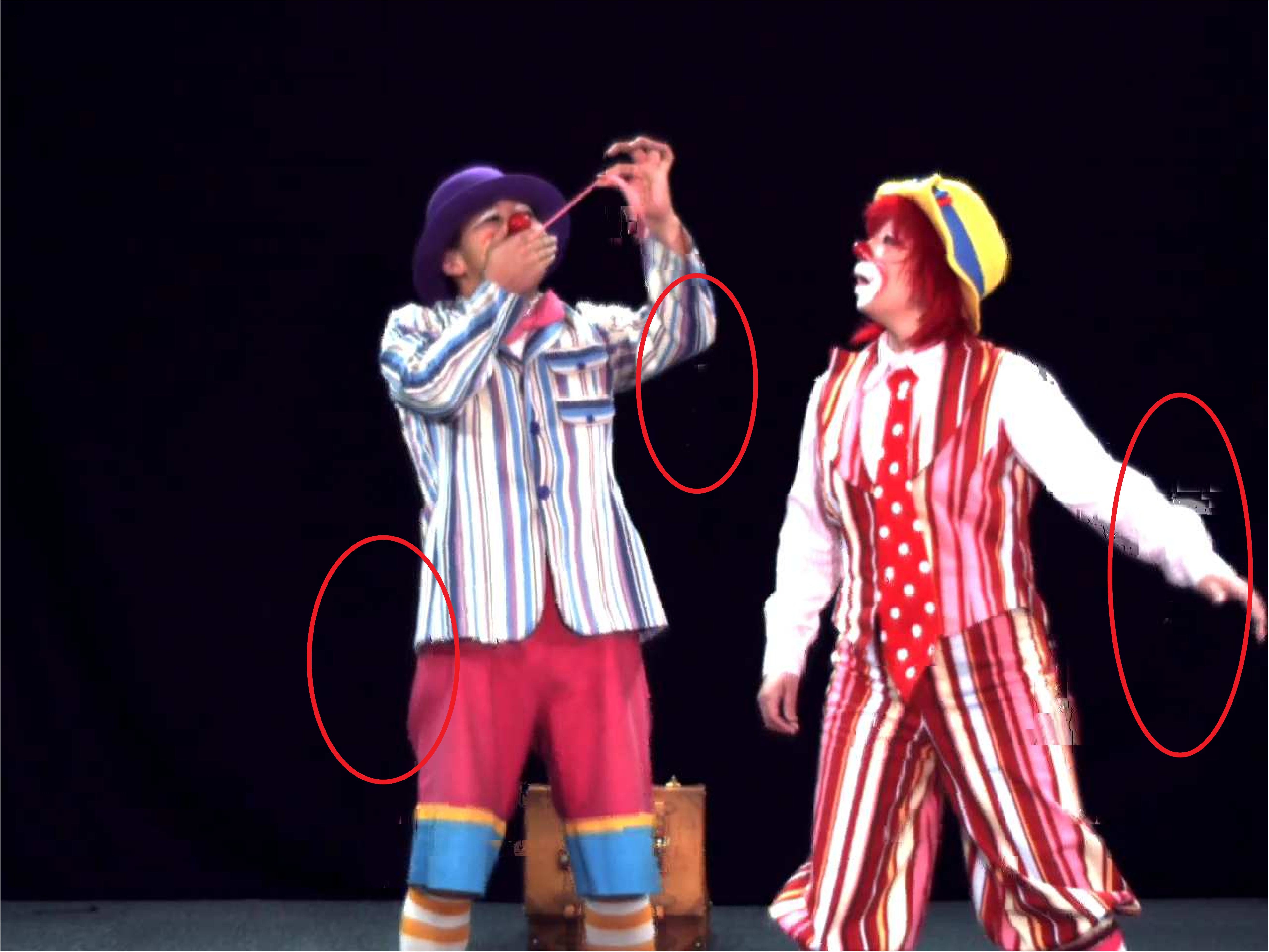} }
\vspace{-2mm}
\caption{Synthesized views for \texttt{Pantomime}, frame $118$, at $5\%$ packet loss. (a) RFC, (b) RPS1 and (c) ARPS schemes.}
\label{res:fig2}
\end{figure}

\begin{figure}[!htbp]
\centering
\subfigure[][]{\includegraphics[width=0.49\textwidth]{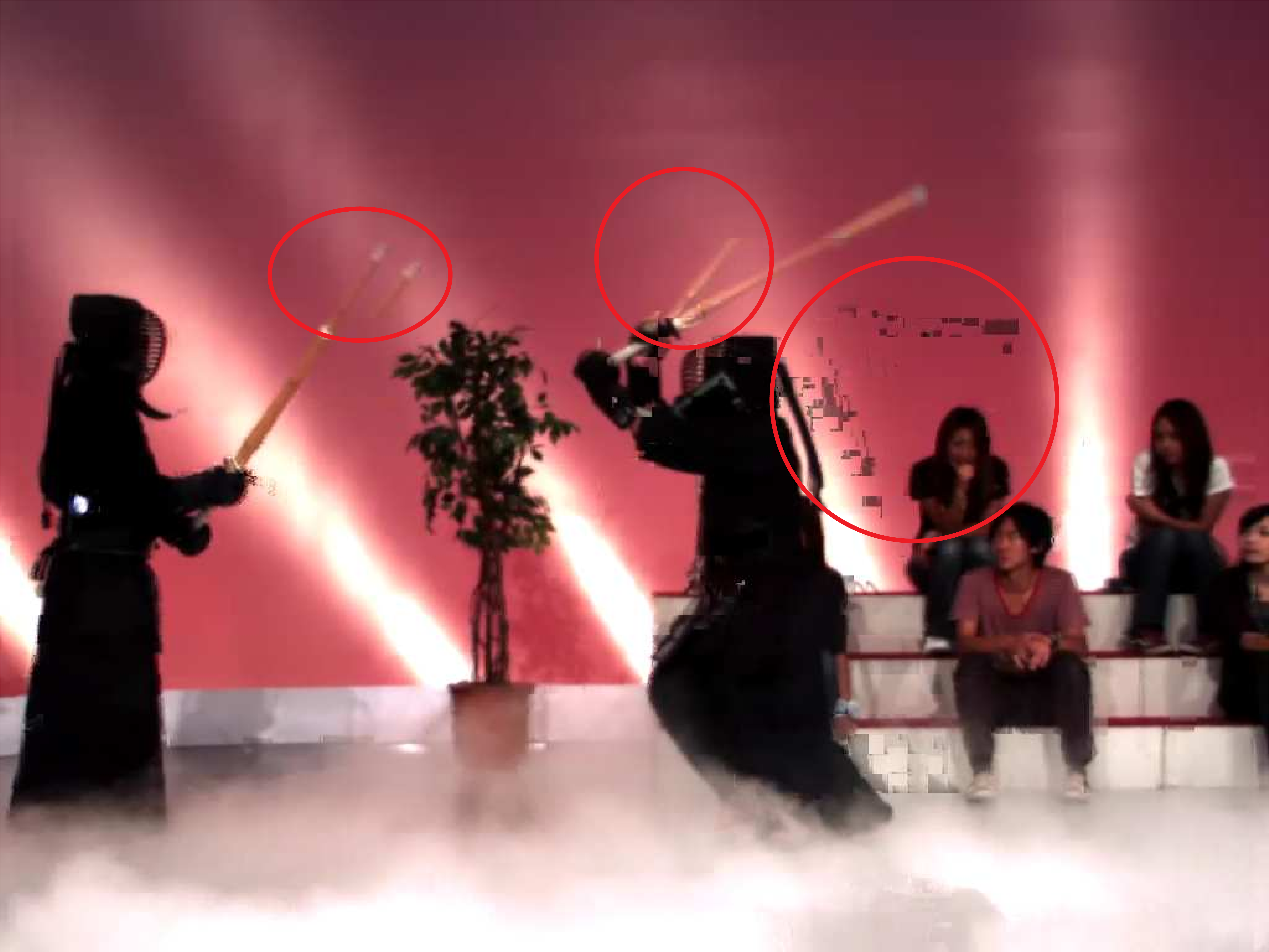}}\\
\subfigure[][]{\includegraphics[width=0.49\textwidth]{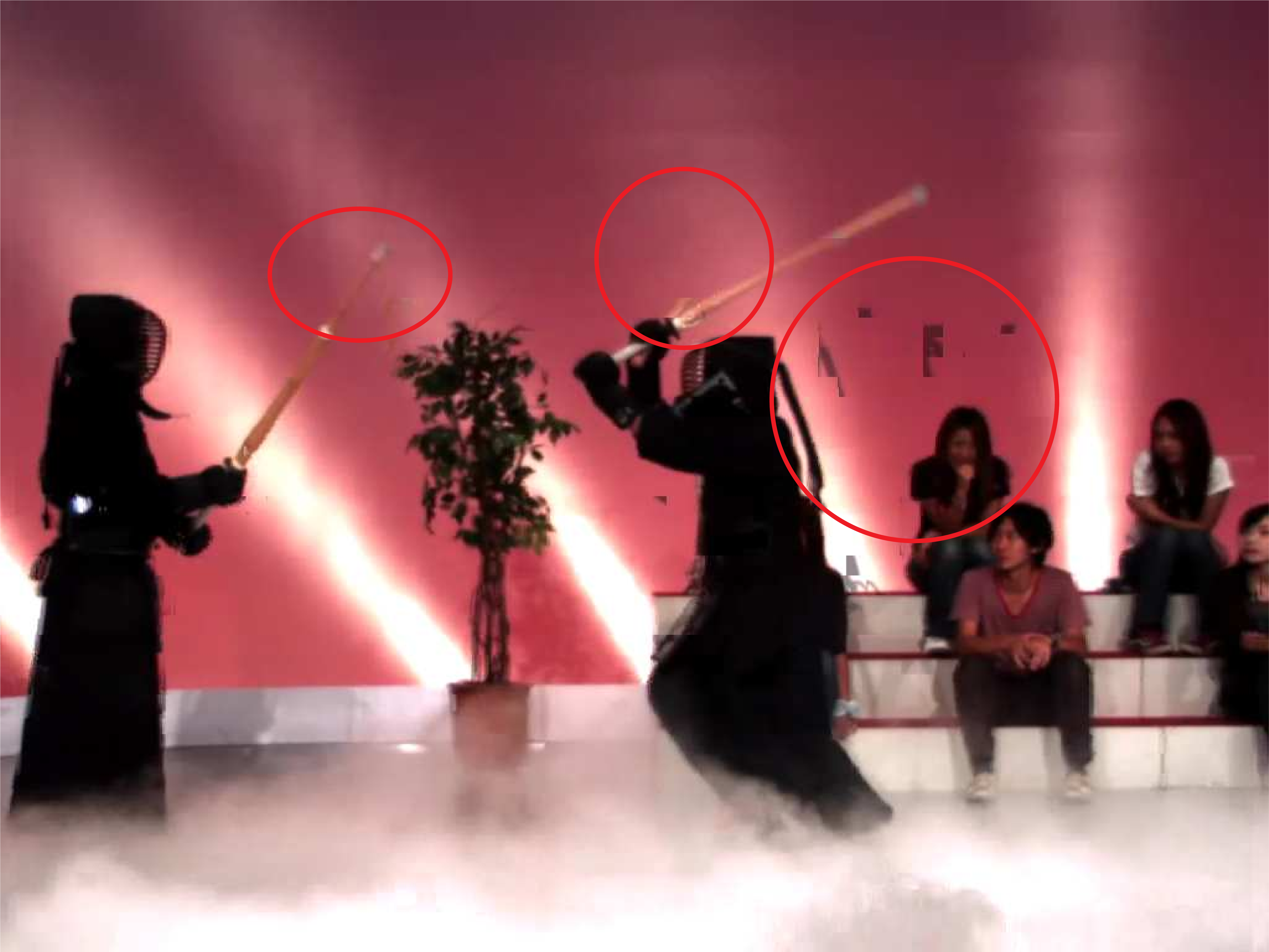}}\\
\subfigure[][]{\includegraphics[width=0.49\textwidth]{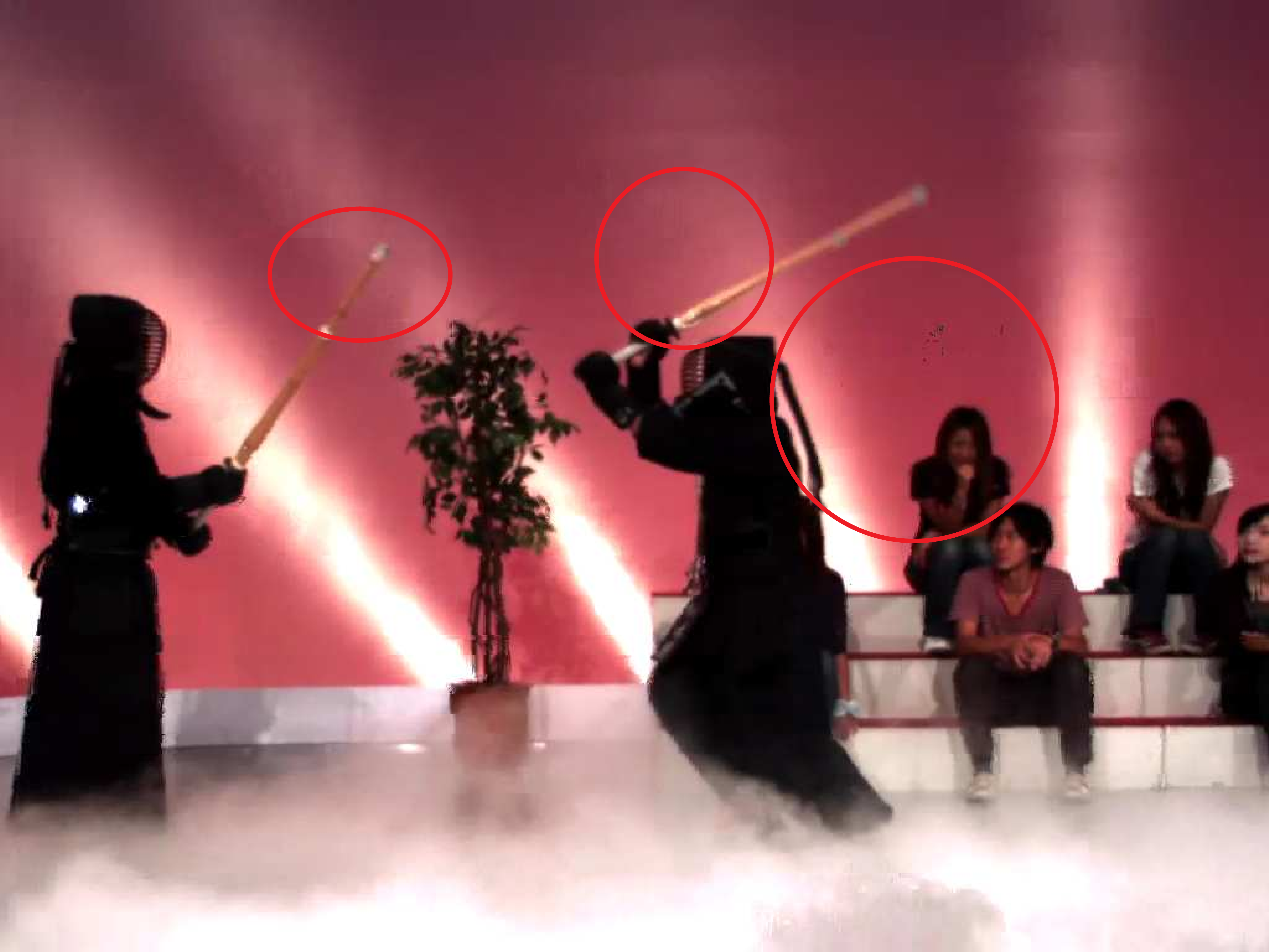} }
\vspace{-2mm}
\caption{Synthesized views for \texttt{Kendo}, frame $22$, at $5\%$ packet loss. (a) RFC, (b) RPS1 and (c) ARPS schemes.}
\label{res:fig3}
\end{figure}

\begin{figure}[!htbp]
\centering
\subfigure[][]{\includegraphics[width=0.49\textwidth]{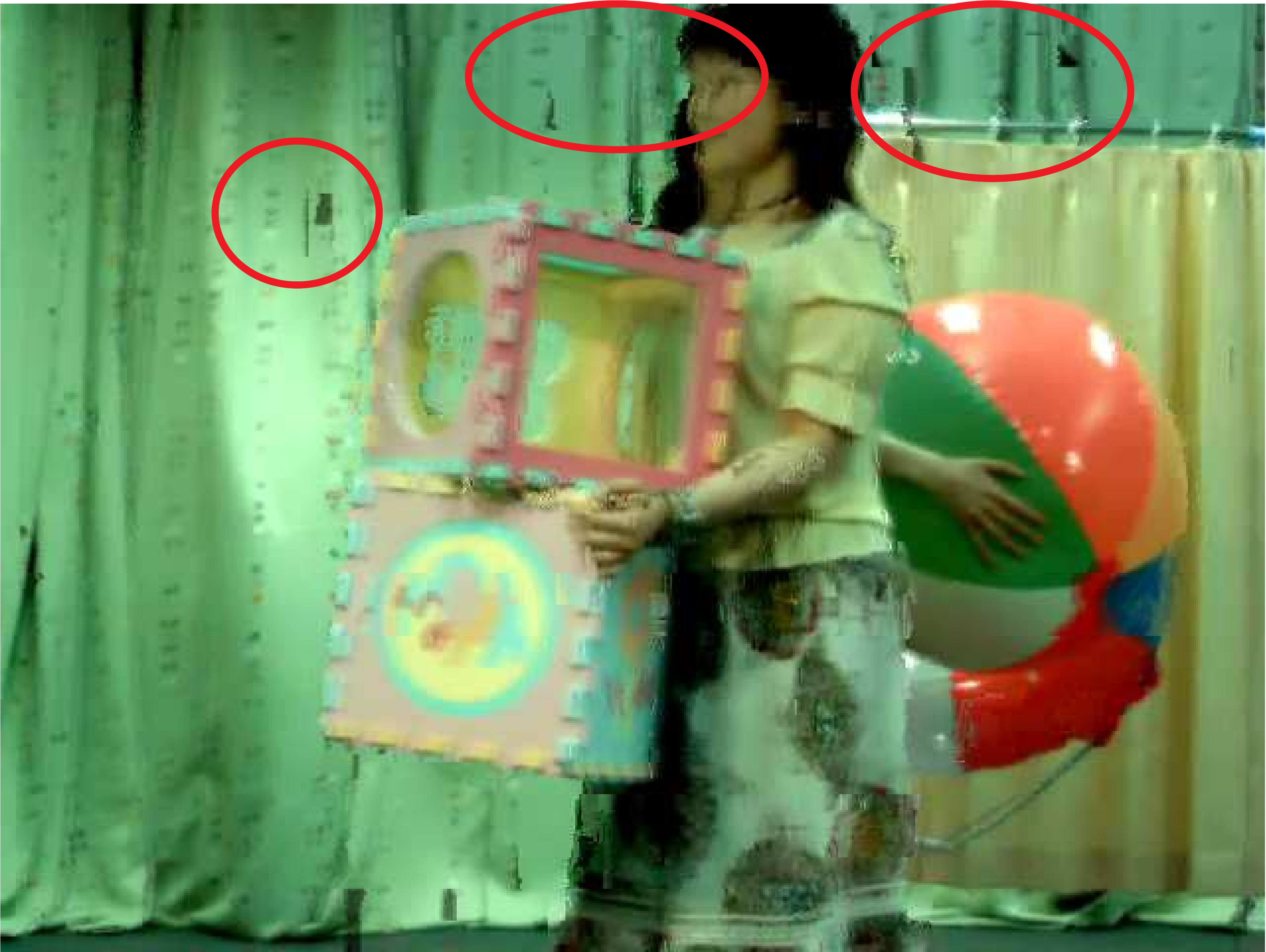}}\\
\subfigure[][]{\includegraphics[width=0.49\textwidth]{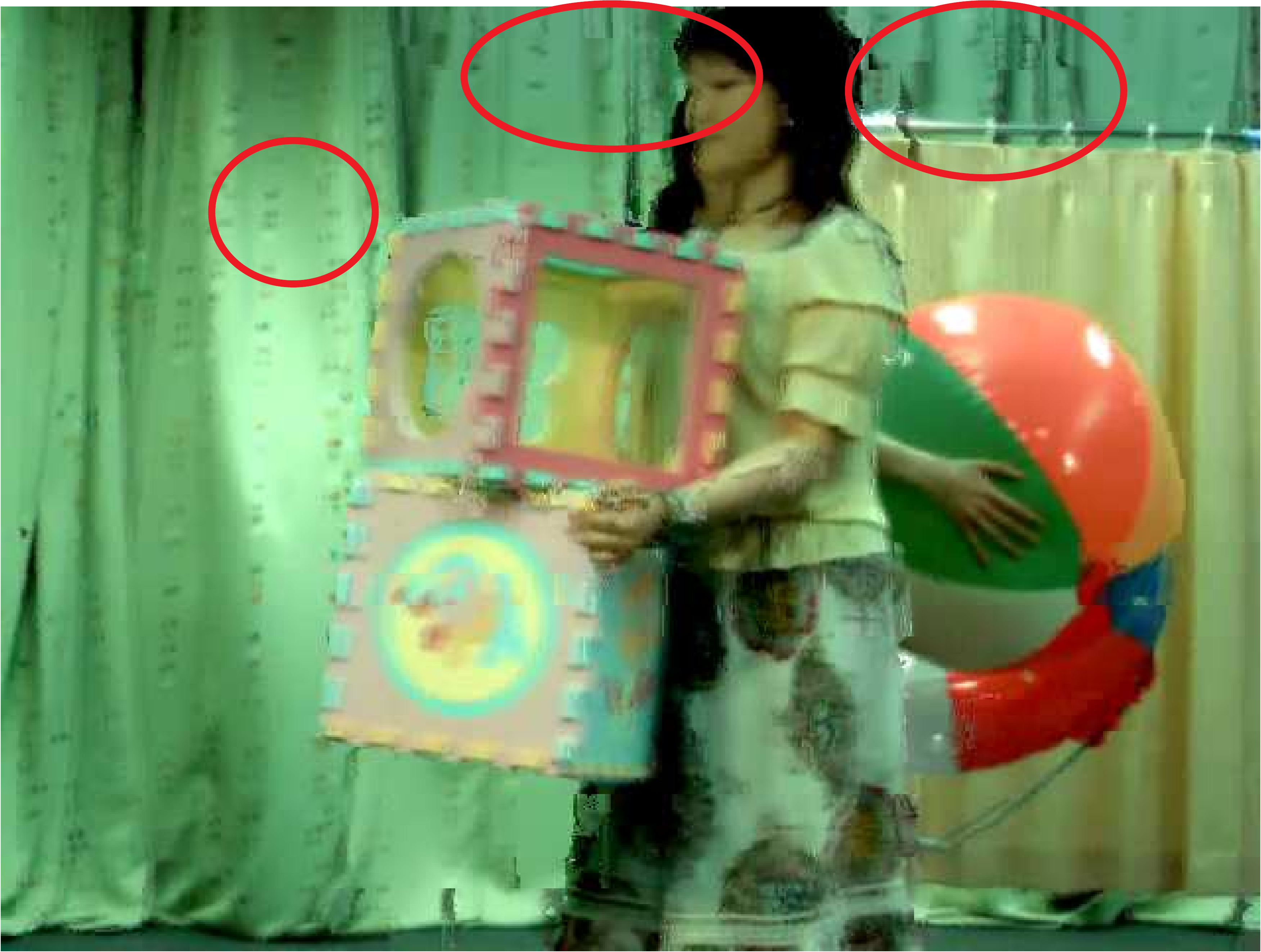}}\\
\subfigure[][]{\includegraphics[width=0.49\textwidth]{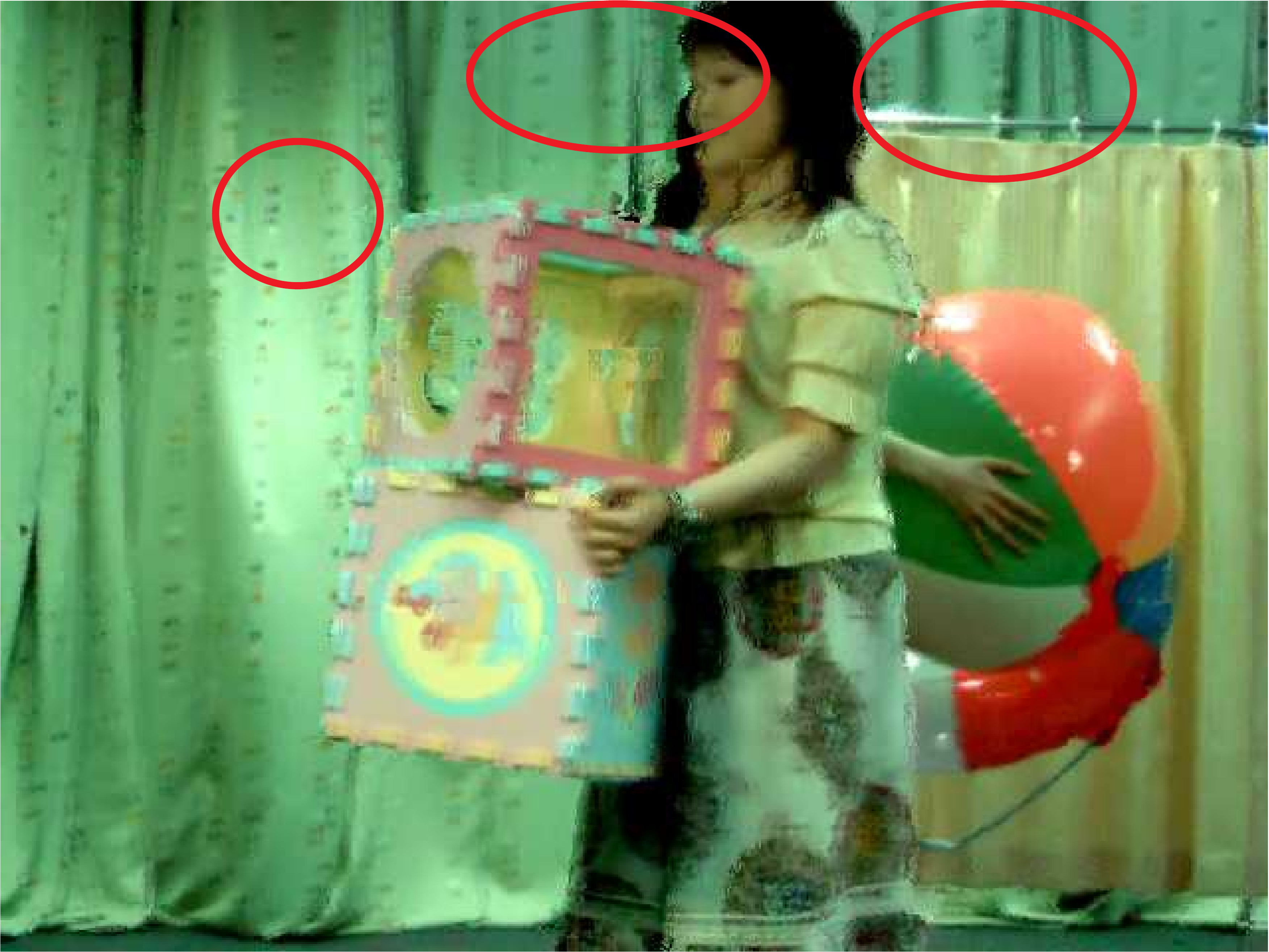}}
\vspace{-2mm}
\caption{Synthesized views for \texttt{Akko and Kayo}, frame $38$, at $5\%$ packet loss. (a) RFC, (b) RPS1 and (c) ARPS schemes.}
\label{res:fig4}
\end{figure}

For subjective comparisons, Figs.~\ref{res:fig2}-\ref{res:fig4} present, for each sequence, synthesized views resulting from the application of
methods RFC, RPS1 and ARPS. The more significant differences among images are circled in red. Visual gains between methods are very significant for both the \texttt{Kendo} and \texttt{Pantomime} sequences and they can be easily perceived at
normal frame rate $30$ fps. 
 
In \texttt{Kendo}, for example, we can see errors in the RFC image (top frame of Fig.~\ref{res:fig3}) around both
swords and above the audience. The errors in the background above the audience are residue created 
by using co-located blocks to conceal errors in past frames, which were not yet 
acknowledged by the feedback channel. The errors around the
swords are most likely caused by packets that were lost in the depth maps, generating
an erroneous horizontally shift. The RPS1 scheme reduces the errors in the background 
and the unequal protection in the depth maps helps to conserve the sharp edges around the swords, 
which avoids such visual degradations, as can be seen in the middle frame of Fig.~\ref{res:fig3}. The proposed adaptive blending 
significantly improves the synthesized view by applying weights that consider the estimated
error in each pixel. The residue errors above the audience is minimal in the bottom frame of Fig.~\ref{res:fig3},  
resulting in a more pleasing subjective viewing. Note that these highly visible difference in visual
quality need not give rise to large PSNR difference, as the artifacts affect only a small portion of the image.

Similar results can be seen in Fig.~\ref{res:fig2} for \texttt{Pantomime} below the elbow and near the waist of the 
clown in the left and in the left arm of the clown in the right. For \texttt{Akko and Kayo} the subjective gains
are not as significant as in the other two sequences and are harder to perceive by
a untrained viewer at regular frame rate. However, it is important to notice that the view synthesis software 
does not perform as well in this sequence as in the other two, therefore it limits the subjective
gains that our adaptive blending can achieve. Nevertheless, the improvement of a particular frame can
be significant as shown in Fig. \ref{res:fig4}.

\section{Conclusion}
\label{sec:conclude}

To enable free-viewpoint video conferencing, in this paper we study a
real-time streaming system where texture and depth videos from two
captured viewpoints are transmitted, so that synthesis and display
of any intermediate viewpoint at receiver is enabled via depth-image-based 
rendering (DIBR). To provide resiliency over loss-prone transmission
networks, we propose first to adaptively blend a synthesized pixel
at receiver, so that the pixels from the more reliable transmitted
view is weighted heavier during synthesis. We then propose a
reference picture selection (RPS) scheme at sender, so that pro-actively
important code blocks containing pixels vital to synthesized view
quality are predicted from older past frames, lowering their expected
error due to error propagation in differentially coded video.
Finally, we analyze synthesized view distortion sensitivities to 
texture versus depth errors, so that relative importance can be
determined for texture and depth code blocks for system-wide RPS
optimization. Experimental results show that our proposed scheme, 
combined with feedback information, can significantly 
outperform a reactive feedback channel, not only
by objectives metrics, but subjectively as well.

\end{document}